\documentclass[nofootinbib,oneside,showpacs,aps,prd,superscriptaddress,amsmath,amssymb,floatfix,11pt]{revtex4}
\usepackage{graphicx}
\usepackage{bm}
\usepackage{color}
\usepackage{multirow}
\usepackage{bookmark}

\usepackage{xcolor}
\hypersetup{
    colorlinks,
    linkcolor={red!60!black},
    citecolor={blue!50!black},
    urlcolor={blue!80!black}
}

\begin{document}
\title{The prospect of charm quark magnetic moment determination}
\date{10 September 2019}							
\author{{A.S.~Fomin}}
\email[]{alex.fomin@cern.ch}
\affiliation{LAL (Laboratoire de l'Acc\'el\'erateur Lin\'eaire), Universit\'e Paris-Sud/IN2P3, Orsay, France}
\affiliation{NSC Kharkiv Institute of Physics and Technology, 61108 Kharkiv, Ukraine}
\affiliation{CERN, European Organization for Nuclear Research, Geneva 23, CH-1211 Switzerland}
\author{S.~Barsuk}
\affiliation{LAL (Laboratoire de l'Acc\'el\'erateur Lin\'eaire), Universit\'e Paris-Sud/IN2P3, Orsay, France}
\author{A.Yu.~Korchin}
\email[]{korchin@kipt.kharkov.ua}
\affiliation{NSC Kharkiv Institute of Physics and Technology, 61108 Kharkiv, Ukraine}
\affiliation{V.N.~Karazin Kharkiv National University, 61022 Kharkiv, Ukraine}
\author{V.A.~Kovalchuk}
\affiliation{NSC Kharkiv Institute of Physics and Technology, 61108 Kharkiv, Ukraine}
\affiliation{V.N.~Karazin Kharkiv National University, 61022 Kharkiv, Ukraine}
\author{E.~Kou}
\email[]{kou@lal.in2p3.fr}
\affiliation{LAL (Laboratoire de l'Acc\'el\'erateur Lin\'eaire), Universit\'e Paris-Sud/IN2P3, Orsay, France}
\author{M.~Liul}
\affiliation{LAL (Laboratoire de l'Acc\'el\'erateur Lin\'eaire), Universit\'e Paris-Sud/IN2P3, Orsay, France}
\affiliation{V.N.~Karazin Kharkiv National University, 61022 Kharkiv, Ukraine}
\author{A.~Natochii}
\affiliation{LAL (Laboratoire de l'Acc\'el\'erateur Lin\'eaire), Universit\'e Paris-Sud/IN2P3, Orsay, France}
\affiliation{Taras Shevchenko National University of Kyiv, 01601 Kyiv, Ukraine}
\author{E.~Niel}
\affiliation{LAL (Laboratoire de l'Acc\'el\'erateur Lin\'eaire), Universit\'e Paris-Sud/IN2P3, Orsay, France}
\author{P.~Robbe}
\affiliation{LAL (Laboratoire de l'Acc\'el\'erateur Lin\'eaire), Universit\'e Paris-Sud/IN2P3, Orsay, France}
\author{A.~Stocchi}
\email[]{stocchi@lal.in2p3.fr}
\affiliation{LAL (Laboratoire de l'Acc\'el\'erateur Lin\'eaire), Universit\'e Paris-Sud/IN2P3, Orsay, France}

\begin{abstract}

In this paper, we discuss the theoretical framework and the experimental measurements of the magnetic moment of the charm baryons.
The $\Lambda_c^+$ magnetic moment is particularly interesting since it is equal to the magnetic moment of the charm quark.
The measurements of the magnetic moments of other charm baryons, such as $\Xi_c$, allow to perform detailed spectroscopy studies.
The magnetic moment of the $\Lambda_c$ can be determined using radiative charmonium decay and the present results
show a tension with majority of theoretical predictions.
As recently pointed out, the magnetic moment of the charm baryons can be directly measured using
bent-crystal experiments at LHC. The possibility of precisely measure the magnetic moments of charm
baryons needs precise measurement of their polarisation and weak decay parameters.
In this paper, we revisit the formalism of the angular analysis needed for these measurements 
and make a detailed evaluation of initial polarisation of deflected $\Lambda_c$ baryons as a function  of crystal orientation.
We found a special orientation of the crystal that gives the opportunity to measure the $\Lambda_c$ dimensionless electric dipole moment almost with the same precision as its $g$-factor, which is more than an order of magnitude more efficient than suggested before.
In conclusion, we stress the importance to perform precise measurements of initial polarisation and weak decay parameters of $\Lambda_c$ baryon to effectively compare the direct and from decay measurements of magnetic moments.

\end{abstract}


\maketitle
\newpage

\tableofcontents

\vspace{.5cm}

\setcounter{footnote}{0}

\section{The magnetic dipole moment of charmed baryon}

The spin 1/2 particles, such as leptons, proton, quarks, have intrinsic magnetic dipole moment (MDM), of the form:
\begin{equation}
{\mu}=\frac{g}{2}\frac{e Q}{2m}
\label{eq:1_v2}
\end{equation}
{where $Q$ is the electric charge in units of the positron charge $e$, and $m$ is the mass of a 
particle\footnote{We use natural units where $\hbar = c =1$.}.  }
Factor $g$ is called {gyromagnetic factor, or $g$-factor},
which is 2 at the classical level while the quantum effect can modify this value.
As this deviation, {anomalous MDM $\kappa \equiv (g -2)/2 $},
comes from the loop effect, it is known to be sensitive to the contributions from new physics:
a heavy particle from new physics can propagate in the loop.

The MDM of electron and muon are one of the most precisely measured quantities in particle physics:
$g_e=2.002 319 304 361 82 (52), g_\mu =2.002 331 841 8 (13)$. The theoretical predictions for these quantity are also
computed at a very high accuracy, e.g. the 5 loop in QED. Intriguingly, a deviation ($3.6 \sigma$ level as of today) is
observed in the muon anomalous MDM, which is one of most significant hints of new physics observed today.

The proton MDM is also measured very precisely, $g_p=5.585 694 702 (17)$. This value is obtained by
using the proton charge and mass in Eq.~(\ref{eq:1_v2}), i.e. $Q=1$ and $m=m_p$. This value being far from 2
is an indication of the proton substructure.

Thus, let us take into account the fact that proton is made of three quarks.
Within the quark model description, the MDM of proton can be computed as a sum of
the MDMs of the three constituent quarks. It is important here to take into account the spin
configuration of the three quarks.
As a result one finds 
\begin{equation}
\mu_p=\frac{1}{3}(4\mu_u-\mu_d)
\end{equation}
where $\mu_{u, d}$  is the MDM of up and down quarks given as
\begin{equation}
{\mu_{q}=\frac{g_q}{2}\frac{e Q_q}{2 m_{q}}},
\end{equation}
{where  $Q_{q}$ and $m_{q}$ are the electric charge and the mass,  and $g_q$ is the $g$-factor of the quark $q = (u, \, d)$.}

In the isospin symmetry limit, $m_u=m_d\equiv m_q$ {and $g_u=g_d \equiv g_q$}, we find
\begin{equation}
{\mu_p=\frac{g_q}{2}\frac{e}{2 m_q}}
\label{eq:protonMDM}
\end{equation}
One can immediately recognize that the result suffers from the uncertainty coming from the quark mass.
In the quark model, the constituent quark mass $m_q=m_p/3$ can be used.
Using this, a comparison of Eq.~(\ref{eq:1_v2}) and Eq.~(\ref{eq:protonMDM})
leads to the value $g_q = g_p/3 \simeq 1.862$.
This result being close to 2 indicates that {the light quarks $u, \, d$ have} little substructure.

The difficulties for concluding whether {the quark} {$g$-factor}
is SM-like or not are three-fold:

(i) the result heavily depends on the quark mass, {in fact from the experiment one can only determine the ratio
$g_q / m_q= g_p / m_p \simeq 5.95$ GeV$^{-1}$;}

(ii) different from the lepton case, the anomalous MDM is induced
by the strong interaction and it can be very large and also scale dependent;

(iii) it has been claimed that the spin of the proton is not carried by quarks but mainly by gluons.

{We note by passing, that the argument (i) can be reversed:
by going to the classical limit $g_q=2$ {one can determine the $u$ or $d$ quark mass},
$m_q= 0.336$ GeV.}

Having said this, the agreement of {this experimental result does } not seem to be just an accident.
The computation of the MDM of neutron in the same model leads to
\begin{equation}
\mu_n= \frac{1}{3}(4\mu_d-\mu_u) = - \frac{g_q}{2} \frac{e}{3 m_q}=-\frac{2}{3}\mu_p,
\label{eq:mu_p_vs_mu_n}
\end{equation}
where the relation between $\mu_n$ and $\mu_p$ is quark mass {and $g$-factor} independent.
It is very well satisfied in the experiment.

Now moving to charmed baryon MDM, the theoretical prediction
suffers from the similar problem as for the proton: the ambiguity due to the charm quark mass dependence.
On the other hand, at the heavy quark limit (i.e. charm quark to be infinitely heavy), it can be derived
that the spin of the charmed baryon is mainly carried by the spin of the charm quark,
which can overcome the problem we encountered in the proton case.

Let us first compute the MDM of $\Lambda_c^+$ in the constituent quark model as before,
i.e. $\Lambda_c^+$ MDM is sum of the MDMs of up, down and charm quarks
in the configuration antisymmetric in spin of the light quarks (see Appendix A).
In this case the $\Lambda_c^+$ MDM is equal to the charm quark MDM:
\begin{equation}
\mu_{\Lambda_c} = \mu_c  = \frac{g_c}{2} \frac{e Q_c}{2 m_c}
\end{equation}
Using the constituent quark mass $m_c=m_{\Lambda_c}-2m_q(1/3\ {\rm GeV})=1.7$ GeV we find
\begin{equation}
\mu_{\Lambda_c}=0.37 \, \frac{g_c}{2} \, \mu_N,
\end{equation}
where the nuclear magneton is $\mu_N =\frac{e }{2m_p }$.
It is curious that the $g$-factor of $\Lambda_c^+$, which is defined
via $\mu_{\Lambda_c}=\tfrac{g_{\Lambda_c}}{2}\tfrac{e}{2m_{\Lambda_c}}$,
is actually close to the charm quark $g$-factor, i.e.
$g_{\Lambda_c}=\tfrac{Q_c \, m_{\Lambda_c}}{m_c}\, g_c\simeq 0.9\, g_c$,  although $\Lambda_c^+$ has a substructure.

There are various models to compute the MDM beyond the quark model.
For example, the so-called Heavy Hadron Chiral Perturbation Theory (HHCPT) is developed~\cite{Wise:1992hn,Burdman:1992gh,Yan:1992gz,Cho:1992gg},
which combines the heavy quark effective theory and the chiral perturbation theory of light hadrons.
It allows to improve theoretical prediction in a systematic manner.

The next to leading order Lagrangian for the MDM of triplet and sextet baryons has been
given in~\cite{Cho:1992nt,Cho:1994vg}.
At the order ${\mathcal{O}}(1/m_Q)$ ($m_Q$ is the heavy quark mass), we have two extra contributions.
First, it is the heavy quark MDM, i.e. the interaction of the photons and the heavy quark constituent inside
of the hadrons, which also induces M1 transition. This term induces the contribution of the quark model.
The second contribution is the photon interaction with the light ``brown mock'' inside of the heavy hadrons.
However, the baryon $\Lambda_c^+$, whose light degrees of freedom are in the spinless state, does not receive contribution to the MDM from this interaction.
As a result, even at this order, the quark model limit results given above hold.
The lack of contributions from the light degrees of freedom seems to be generic and theoretical predictions using different models show
that the MDM of $\Lambda_c$ is close to the one predicted by the constituent quark model.

MDM predictions using various theoretical models can
be summarized as~\cite{Franklin:1981rc,Barik:1984tq,Savage:1994zw,SilvestreBrac:1996bg,Aliev:2001ig,JuliaDiaz:2004vh,Kumar:2005ei,Faessler:2006ft,Karliner:2006ny,Majethiya:2008fe,Patel:2007gx,Sharma:2010vv,Bernotas:2013}:
\begin{equation}
\mu_{\Lambda_c} = (0.34 - 0.43) \, \mu_N, \label{eq:LcMMth-v2}
\end{equation}
although there are exceptions falling out of these bounds.
In particular, Ref.~\cite{Zhu:1997as}, using the QCD spectral sum rule approach, gives $\mu_{\Lambda_c} =( 0.15\pm0.05)\, \mu_N$, while Ref.~\cite{JuliaDiaz:2004vh} in the Dirac point-form dynamics obtains $\mu_{\Lambda_c} = 0.52 \, \mu_N$, and Ref.~\cite{Wang:2018xoc} in the next-to-next-to-leading order in
the HHCPT gives $\mu_{\Lambda_c} = (0.24\pm0.02) \, \mu_N$.
It should be noted that each theoretical model fits the charm quark mass with various observables.
In this sense, the charm quark mass uncertainty is included in this value.

\section{The prediction of the magnetic dipole moment of $\Lambda_c$ using the radiative charmonium decays}

It turned out that the charm quark MDM is most precisely measured by the quarkonium radiative decays as of today. In this section, using this result, we estimate the $\Lambda_c$ MDM. 

The process used by the CLEO~\cite{Artuso:2009aa} and BESIII~\cite{Ablikim:2017gom} collaboration is  the cascade radiative decay, $\psi(3686) \to \gamma_1 \chi_{c1,2}$ followed by $\chi_{c1,2} \to \gamma_2 J/\psi(\to l^+l^-)$, where the initial $\psi(3686) $ is produced from the $e^+e^-$ collision. 

Using the non-relativistic model, the cascade radiative decay, $\psi(3686) \to \gamma_1 \chi_{c1,2}$ followed 
by $\chi_{c1,2} \to \gamma_2 J/\psi(\to l^+l^-)$ is computed (see \cite{Karl:1975jp, Karl:1980wm} for detail). 
We have 5 observable angles: $\chi_{c1,2}$  direction with respect to $e^+e^-$ in the centre of mass system ($\theta_0$), 
the angle between $\gamma_1$ and $\gamma_2$ in the rest frame of $\chi_{c1,2}$ ($\theta_1, \phi_1$) 
and angle between $l^+$ and $\gamma_2$ in the rest frame of $\chi_{c1,2}$ ($\theta_2, \phi_2$). 
The helicity amplitudes of  $\psi(3686) \to \gamma_1 \chi_{c1,2}$ and $\chi_{c1,2} \to \gamma_2 J/\psi$ 
decays are parameterised by the coefficients $a_i^J$ and $b^J_i$, respectively. 

The normalised M2 contributions, $b^{1,2}_2$ and $a^{1,2}_2$ from the $\psi(3686) \to \gamma_1 \chi_{c1,2}$ and $\chi_{c1,2} \to \gamma_2 J/\psi$, respectively, are related to the mass of the charm quark $m_c$ and its anomalous magnetic moment $\kappa$ (see \cite{Artuso:2009aa} for detail). 
In the ratios $\beta = b^1_2 / b^2_2$ and $\alpha = a^1_2 / a^2_2$, the $m_c$ and $\kappa$ cancel to first order in $E_\gamma / m_c$. The ratios thus receive clear numerical predictions of $\beta = 1.000 \pm 0.015$ and \mbox{$\alpha = 0.676 \pm 0.071$}, respectively~\cite{Artuso:2009aa}. Recently, the BES~III experiment reported~\cite{Ablikim:2017gom} the measurement of the M2 amplitudes and the determination of the two ratios $\beta = 1.35 \pm 0.72$ and $\alpha = 0.617 \pm 0.083$, in agreement with the theory prediction. 

The precision of the $b^{1,2}_2$ and $a^{1,2}_2$ measurements reported by BES~III is dominated by the available statistical sample, and is expected to be improved by future experiments with larger collected data samples. Among important systematic uncertainties are photon detection, efficiency estimates with the simulation assuming the phase space, kinematic fit and fitting technique. With improved electromagnetic calorimeter and the efficiency determined in bins of relevant angular variables, systematic uncertainty is also expected to be significantly improved in the next-generation experiments. 
In the BES~III analysis~\cite{Ablikim:2017gom}, the $(1 + \kappa_c)$ measurement, which can be related to $g_c$ by 
\begin{equation}
{1+\kappa_c=\frac{g_c}{2}}
\end{equation} 
was performed: 
\begin{equation}
\frac{g_c}{2}=-\frac{4m_c}{E_{\gamma_2}}a^1_2=1.140\pm 0.051\pm 0.053\pm 0.229
\label{eq:BESIIIMC}
\end{equation}
where the last systematic error is coming from the charm quark mass ambiguity {$m_c = 1.5 \pm 0.3 $ GeV.} 

What would be the implication of this result? Indeed, the obtained value of $g_c$ is close to 2 but the precision is 
limited by  the uncertainty from the charm quark mass. In fact, the charm quark would receive a radiative correction, 
from the strong interaction, which would also induce uncertainty. As in the case of the muon anomalous 
magnetic moment, there is a chance that the charm quark anomalous magnetic moment is non-SM like. 
However, the SM prediction of the $g_c$ contains an ambiguity as a concept. 
This problem can be solved only when we chose a theoretical model which allows to consistently 
calculate the charm quark anomalous magnetic moment effect inside of hadrons.
In the following we prefer to write the result in Eq.~(\ref{eq:BESIIIMC}) in terms of ratio between $g_c$ and the charm mass quarks as
\begin{equation}
\frac{g_c}{2 m_c} =0.76 \pm 0.05 \, \,  {{\rm GeV}^{-1}} 
\label{eq:BESIIIMC-bis}
\end{equation}

Since the magnetic moment of charm quark is proportional to $g_c/{2 m_c}$, 
the experimental results given in Eq.~(\ref{eq:BESIIIMC}), can provide a prediction of 
the  $\Lambda_c$ magnetic moment {in the constituent quark model}  
{\it without any charm quark uncertainty}
\begin{equation}
 \mu_{\Lambda_c} =   \mu_c = \frac{g_c}{2 m_c}  \frac{2}{3} m_p \, \mu_N  =   ( 0.48 \pm 0.03)\, \mu_N.
\label{eq:tension}
\end{equation}

If we compare this with the theoretical predictions in the end of the previous Section, we can conclude that there is a tension with the majority of theoretical predictions. 
In particular, the deviation with calculation \cite{Zhu:1997as} is 5.7 $\sigma$, with the NNLO HHCPT \cite{Wang:2018xoc} the deviation is 6.7 $\sigma$. 
On the other hand, there are theoretical models which do not contradict to Eq.~(\ref{eq:tension});  
for example, the calculation in \cite{Aliev:2001ig} agrees with the value in Eq.~(\ref{eq:tension}) on the level of 1.4 $\sigma$.

In order to increase the significance of this discrepancy and to observe a possible new physics contribution, 
what would be needed are 
i) to achieve a better precision of the measurement given in~(\ref{eq:BESIIIMC-bis}) by further improving radiative charmonium decay at BESIII and a possible future charm factory, 
ii) to achieve a direct measurement of $\Lambda_c$ magnetic moment at a equivalent 
precision.
We will briefly discuss on i) in the following while ii) will be discussed in the section 5 and 6. 
In both cases we should aim to have an experimental precision at 5$\%$ or better.

Theory calculations of $b^{0,1,2}_2$ and $a^{0,1,2}_2$ to the next order in $E_\gamma / m_c$ are therefore of primary importance. A dependence of the corrections on $E_\gamma$ and $m_c$ is expected to be different, so that the experimental determination of different amplitude will provide truly complementary information. This can be possible by BESIII and also at the future tau-charm factories.

Another path can be the measurement of the absolute values of $b^{1,2}_2$ and $a^{1,2}_2$ instead of the ratios. While part of the systematic uncertainties will not cancel for absolute measurements, the system can be over-constrained to verify model assumptions. Comparison of the all four values of $b^{1,2}_2$ and $a^{1,2}_2$ measured experimentally to theory predictions will provide complementary information. Taking into account correlations of experimental measurements and retaining only variable that yield identical theoretical interpretations, the extracted values for $\kappa$ or $\kappa \oplus m_c$ can be averaged. Involving measurements with intermediate $\chi_{c0}$ and $\eta_c (2S)$ states would allow a simultaneous fit to the $m_c$ and $\kappa$ variables to be performed using the eight quasi-independent measurements.

Higher-order multipole amplitudes can be extracted from the angular distributions of the final-state particles. They were first considered in Ref.~\cite{Ablikim:2017gom} by the BES~III experiment, who performed a simultaneous unbinned maximum likelihood fit according to the procedure from Refs.~\cite{Karl:1975jp, Karl:1980wm}. The relevant angular variables are the polar angle of $\gamma_1$ with respect to the beam axis, in the $\psi(3686)$ rest frame, $\theta_1$; the polar, $\theta_2$, and azimuthal, $\phi_2$, angles of $\gamma_1$ with respect to the direction of $\gamma_1$, in the $\chi_{cJ}$ rest frame ($\phi_2 = 0$ in the electron-beam direction); the polar, $\theta_3$, and azimuthal, $\phi_3$, angles of $l^+$ from the $J/\psi$ decay with respect to the direction of $\gamma_2$, in the $J/\psi$ rest frame ($\phi_3 = 0$ in the $\gamma_1$ direction).

\section{The charmed baryon spectroscopy and the magnetic dipole moment}

Recently LHCb experiment as well as $e^+e^-$ machine, such as BESIII and Belle II are making a great progresses in the charmed baryon spectroscopy. The first observation of the doubly charmed baryon at LHCb~\cite{Aaij:2017ueg} has triggered various theoretical investigation of charmed baryon weak decays as well. In this section, we show that 
the MDM which reflects the spin configuration of the internal degree of freedom of  baryon can be a powerful tool for identification of charmed baryons. 

Let us first derive the MDM of different charmed baryons. 

First of all, it turned out that the two remaining triplet (spin 1/2, anti-symmetric) charmed baryon, $\Xi_c^0$ and $\Xi_c^+$ have the same MDM as $\Lambda_c$ in the quark model: 
\begin{equation}
\mu_{\Xi^{0,+}_c}= \mu_c =\mu_{\Lambda_c}
\label{eq:XicMDM}
\end{equation}
A confirmation of this relation is  an important role to test the quark model description. 

In~\cite{Savage:1994zw} and ~\cite{Banuls:1999mu}, higher order corrections to this relation are discussed. The reference~\cite{Savage:1994zw} discusses the so-called spin-symmetry breaking effect, which typically induces the $\Sigma_c^*-\Sigma_c$ mass splitting. It comes form a loop diagram with $\Sigma_c^{(*)}$ and $\pi, K$ in the loop. This contribution leads to a sub-leading effect  (order $1/m_Q^2$) to  the MDM of $\Xi_c^{+, 0}$. There are two input parameters, $\Delta m$ and $g_2$\footnote{In~\cite{Savage:1994zw}, it is $g_3$.} but these can be fixed by the $\Sigma_c^*-\Sigma_c$ or $\Xi^{\prime *}_c-\Xi^{\prime}_c$ mass splitting and the charmed baryon strong decays, respectively. In recent years, there are a lot of progresses to determine $g_2$: the latest fit to the experimental dat gives  $0.989^{+0.019}_{-0.042}$~\cite{Cheng:2015naa} and the lattice QCD result shows $0.71\pm 0.13$~\cite{Detmold:2012ge}. Using the former result, to be conservative, we can find that the relation Eq.~\ref{eq:XicMDM} would be modified slightly:  
\begin{equation}
\mu_{\Xi_c^{+/0}}= \mu_{\Lambda_c^+}\pm (0.051\pm 0.001). 
\end{equation}
Note that this result does not depend on the charm quark mass. 
In~\cite{Banuls:1999mu}, the SU(3) breaking effect has been further included. In this case, we can not obtain all the parameters from other experiments, which causes additional theoretical uncertainty. However, this contribution is typically at order $1/m_Q\Lambda_\chi^2$ and can be small. 

For the sextet (spin 1/2, symmetric) charmed baryons, the situation is very different. {We find (Appendix A)}: 
\begin{equation}
\mu_{\Sigma^{++}_c}=-\frac{1}{3}\mu_c+\frac{4}{3}\mu_u, \quad 
\mu_{\Sigma^+_c}=-\frac{1}{3}\mu_c+\frac{2}{3}\mu_u+\frac{2}{3}\mu_d, \quad 
\mu_{\Sigma^0_c}=-\frac{1}{3}\mu_c+\frac{4}{3}\mu_d 
\end{equation}
which leads to the values (with $m_c=m_{\Sigma_c}-2m_q \, (1/3\ {\rm GeV})=1.8$\ GeV) $\mu_{\Sigma_c^{++}}=2.54\, \mu_N$,  
$\mu_{\Sigma_c^+}=0.54 \, \mu_N$, $\mu_{\Sigma_c^0}=-1.46\, \mu_N$. 
Even though this numerical values suffer from the quark mass uncertainty, the sign for $\Sigma_c^0$ seems to be opposite to the one of $\Lambda_c$, and the MDM of doubly-charmed $\Sigma^{++}_c$ is much larger than $\Lambda_c$,  which would be also interesting to be tested.  Note that the main decay channel of $\Sigma_c$ is $\Lambda_c^+\pi$. 

Finally, let us discuss the another sextet (spin 1/2, symmetric) charmed baryon $\Xi_{c}^{\prime +, 0}$. These baryons have the same quark content as $\Xi_c^{+, 0}$ but their wave functions are SU(3) flavour symmetric. Since these states have the same quark contents and the same spin, they can mix with the triplet $\Xi_c^{+, 0}$ states. At the infinite mass limit, though, this mixing is zero, i.e. $\Xi_c^{+, 0}$ is the pure anti-symmetric and $\Xi_{c}^{\prime +, 0}$ is the pure symmetric state. Indeed, two states are observed, one at $\sim2468\,$MeV and the other $\sim2577\,$GeV. The latter decays radiatively to the former. Whether these observed two states (mass eigenstates) are the pure anti-symmetric and symmetric states (flavour eigenstate) is not known though it can offer an excellent test of the heavy quark limit. In the following, we show that the MDM measurement, which are the most sensitive to the flavour symmetry of the constituent quarks,  can be used to answer this question. 

The MDM of  $\Xi_{c}^{\prime +, 0}$ yields: 
\begin{equation}
\mu_{\Xi^{\prime +}_{c}}=-\frac{1}{3}\mu_c+\frac{2}{3}\mu_u+\frac{2}{3}\mu_s, \quad 
\mu_{\Xi^{\prime 0}_{c}}=-\frac{1}{3}\mu_c+\frac{2}{3}\mu_d+\frac{2}{3}\mu_s 
\end{equation}
which leads to  $\mu_{\Xi^{\prime +}_{c}}=\mu_{\Sigma_c^+}, \ \mu_{\Xi^{\prime 0}_{c}}=\mu_{\Sigma_c^0}$ at the SU(3) limit. The theoretical uncertainty might be  larger than the case of $\Sigma_c^{0, +}$ due to $SU(3)$ however, we would still expect the MDM of the $\Xi_{c}^{\prime 0}$ to have an opposite sign comparing to the one of $\Xi_{c}^{0}$. This result implies that the MDM measurements  are very sensitive to resolve the deviation between the flavour and the mass eigenstate of $\Xi_c$'s. In particular, the equality of the MDM of $\Xi_c^{+, 0}$ and $\Lambda_c$ in Eq.~(\ref{eq:XicMDM}) does not depend on the quark masses and the most precise test can be performed. Thus, we will investigate  in the following section, this equality in more detail. 
It should be noted that contrary to the triplet (anti-symmetric) baryons, the sextet (symmetric) baryons receive the next leading order long distance contribution (the photon interacting with light degree of freedom), which are quite sizeable. Neverthelss, most of the theoretical predictions (see e.g.~\cite{Faessler:2006ft}) confirm the negative MDM for $\Xi_c^{\prime 0}$, which can be used to clarify the issue of the $\Xi_c-\Xi_c^{\prime}$ mixing as discussed earlier. 

In summary, a measurement of $\mu_{\Lambda_c}$ as well as $\mu_{\Xi_c}$ at a high precision will be highly appreciated to distinguish different spin configuration of the charmed baryon states.

\section{Magnetic dipole moment measurement of charmed baryons using bent crystal}

The MDMs of baryons containing $u$, $d$ and $s$ quarks have been extensively studied and measured.
The experimental results are all obtained by using the conventional methods, namely the measurement of the precession angle of the polarisation vector when particle is travelling through an intense magnetic field by analysing the angular distribution of the decay products.

No measurement of MDMs of charm or beauty baryons has been performed so far.
A reason of the non-availability of experimental information is because the lifetimes of charm/beauty baryons are too short to measure the MDM by standard techniques.

One proposal to meet the challenge of measuring the MDMs of baryons with heavy flavoured quarks is to use the strong effective magnetic field inside the channels of a bent crystal instead of the conventional magnetic field to process the polarisation vector and measure the MDM.
The detailed precession theory has been developed by~\cite{Baryshevsky:1979,Lyuboshits:1979qw,Kim:1982ry}.

Shortly, in a curved crystal the electrostatic field of the atomic planes deflecting the particle transforms into a magnetic field in the particle's rest frame. Thus the spin precession angle $\phi$ is
\begin{equation}
\phi  =	\omega
		\left(1+\gamma\, \frac{g - 2}{2}\right)
		\ \ {\rm for}
		\ \ \gamma \gg 1,
\label{eq:precession}
\end{equation}
where  $\gamma$ is the Lorentz factor, $g$ is the $g$-factor or dimensionless MDM of baryon, and  $\omega$  is the deflection angle of the channelled particle. 
From a measurement of  $\gamma$, $\phi$ and $\omega$  of the channeled particles, we can determine $g$ and hence the particle's MDM.
We expect that channeling can provide the equivalent magnetic fields up to several hundreds of Tesla, thus offering the potential of significant precession angles even when the length of the bent crystal is of order of cm.

E761 Collaboration (1992) had demonstrated the feasibility of this idea by measuring the MDM of the strange $\Sigma^+$ baryon~\cite{Chen:1992wx,Chen:1992ai} using the decay into p$\pi^0$.

Recently a few papers have appeared~\cite{Baryshevsky:2016cul,Bezshyyko:2017var} proposing experiments to measure the MDM of the $\Lambda_c$ and other charmed charged baryons at LHC top energies.
In~\cite{Fomin:2018ybj} the method for measuring the electromagnetic dipole moments of the $\tau$ lepton using double or triple crystal setups at LHC was proposed.
The clear advantages of the use of LHC are the much larger boost and the possibility of using well-known detectors.
The unavoidable drawback is the complex integration of the crystals into the LHC vacuum pipe in the respect of the machine protection requirements.
However, the recent success of crystal-collimation tests of the UA9 Collaboration~\cite{Scandale:2015gva,Scandale:2016krl}, may provide the necessary technical know-how for such a complex task.  

The experiment foresees the installation of a bended crystal in the halo of the LHC to obtain an intense collimated proton beam.
Polarised heavy baryons are produced by strong interaction of this proton beam impinging into a few mm (tungsten) target.
A large angle bended crystal, located downstream of the target will induce the rotation of the polarisation vector of the heavy baryons.
The change of the polarisation is studied by performing an angular analysis of the decay products of the heavy baryons using either one of the LHC existing detectors or a dedicated new one.

Our goal is to measure the $\Lambda_c$ magnetic moment at {\it a few \%} level. 
As presented  in~\cite{Bezshyyko:2017var}, the sensitivity depends on two factors. 
The first is to have an experimental setup capable to collect enough statistic.
This studies have been made in details using the LHCb detector~\cite{Bagli:2017foe}.
In~\cite{Mirarchi:2019ft} two possible layouts of such a setup are reported, together with a thorough evaluation on their expected performance and impact on LHC operations.
The second factor is to know precisely the initial polarisation of $\Lambda_c$ and to use the most suitable $\Lambda_c$ decay channels giving the greatest sensitivity to the polarisation measurements.
Let us elaborate this second point and discuss on our strategy.

The sensitivity depends on the precision of $\phi$ in Eq.~(\ref{eq:precession}), which represents the spin precession of $\Lambda_c$, ${\rm i.e.}$ the change of the polarisation.
The polarisation of $\Lambda_c$ can be measured, in general, by the angular distribution of its  decay $\Lambda_c\to BP$ ($B$ is a baryon and $P$ is pseudoscalar meson, namely pion or kaon)
\begin{equation}
\frac{1}{N}\frac{dN}{d\cos \theta}=\frac{1}{2}(1+\alpha {\xi} \cos \theta) \label{eq:AD}
\end{equation}
where the $\xi $ is the polarisation projection of $\Lambda_c$,  and the $\theta$ is the angle between $\Lambda_c$ polarisation axis and the final baryon direction $\vec{n}_{\rm baryon}=\vec{p}_{\rm baryon}/|p_{\rm baryon}|$.
The $\alpha$ is called asymmetry parameter, which represents the forward-backward asymmetry of the final state baryon with respect to the initial $\Lambda_c$ polarisation direction.
This asymmetry is non-zero only when the decay is induced by a parity violating interaction.

\subsection{Initial polarisation}

\begin{figure}[t]
\begin{center}
\includegraphics[width=.49\textwidth]{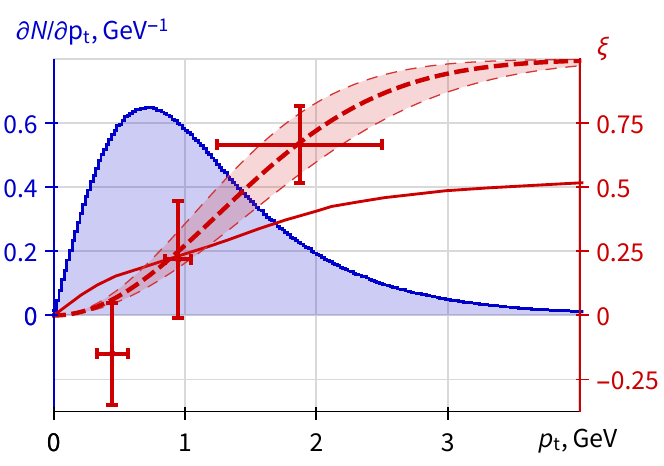}
\includegraphics[width=.49\textwidth]{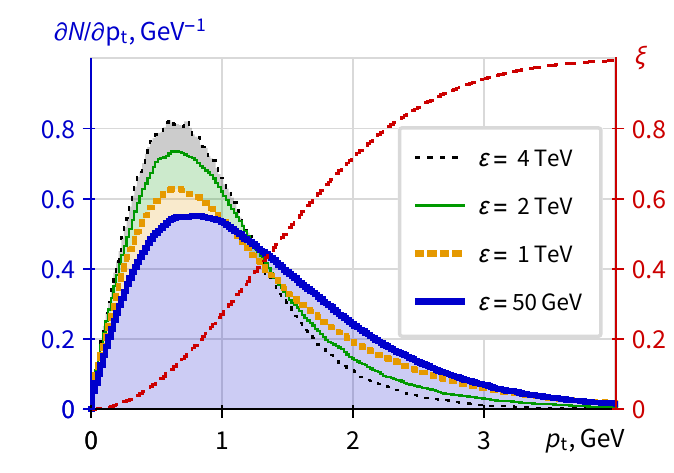}
\end{center}
\caption
{
$\Lambda_c$ initial polarisation as a function of transverse momentum (red curves) on top of its distribution over transverse momentum (histograms).
Red solid curve --- theoretical prediction~\cite{Goldstein:1999hp},
red crosses --- experimental data~\cite{Aitala:1999uq},
red dashed curve --- experimental data fit by equation~\eqref{eq:Xi(pt)};
blue histogram (left plot) --- distribution over transverse momentum of all $\Lambda_c$ produced in a fixed target by $6.5\,$TeV protons,
histograms (right plot) --- same for specific energies of $\Lambda_c$ indicated on the right.
Here the polarisation is projected on the $\vec{p}_{\Lambda_c}\times\vec{p}_{\rm beam}$ direction.
}
\label{fig:P(pt)}
\end{figure}

The experimental data~\cite{Korner:1978tc,Aitala:1999uq} together with theoretical predictions~\cite{Dharmaratna:1996xd,Goldstein:1999jr,Goldstein:1999hp} shows that  $\Lambda_c$ baryons produced in a fixed target are polarised, the polarisation vector is orthogonal to production plane, directed opposite to $\vec{p}_{\rm beam}\times\vec{p}_{\Lambda_c}$, and the absolute value of polarisation grows with transverse momentum (see Fig.~\ref{fig:P(pt)}).
In~\cite{Bezshyyko:2017var} the analysis of this data together with $\Lambda_c$ spectra angular distribution obtained from Pythia~\cite{Sjostrand:2007gs} 
shows that the average absolute value of polarisation of $\Lambda_c$ produced in the fixed target is $|\xi|=0.40(5)$. In the current paper we would like to estimate initial polarisation more accurately.

We extrapolate experimental data with the following expression for polarisation as a function of transverse momentum:
\begin{equation}
|\xi| = 1-{\rm e}^{-\dfrac{p_{\rm t}^2}{\,2\,\langle p_{\rm t}^2\rangle\,}}
\label{eq:Xi(pt)}
\end{equation}
where $\langle p_{\rm t}^2\rangle=1.26(20)\,{\rm GeV}^2$ is a typical transverse momentum of produced $\Lambda_c$ baryons.

The distribution over transverse momentum of  $\Lambda_c$ produced in a fixed target by $6.5\,$TeV protons is obtained using Pythia v.8.240 accounting all soft QCD processes.
Using this data we obtain the root mean square of initial polarisation $\xi^{\rm rms}=0.46(6)$.
We assume that polarisation is a function of transverse momentum and does not depend on  $\Lambda_c$ energy.
On the other hand, the distribution over the transverse momentum varies with $\Lambda_c$ energy (see Fig.~\ref{fig:P(pt)}, right).
Thus, the average polarisation depends on the energy: $\xi^{\rm rms}\approx0.50(6)$ for $\Lambda_c$ energy $\varepsilon=50\,$GeV, and $\xi^{\rm rms}\approx0.34(6)$ for $\varepsilon=4\,$TeV.

\begin{figure}[t]
\begin{center}
\includegraphics[width=.9\textwidth]{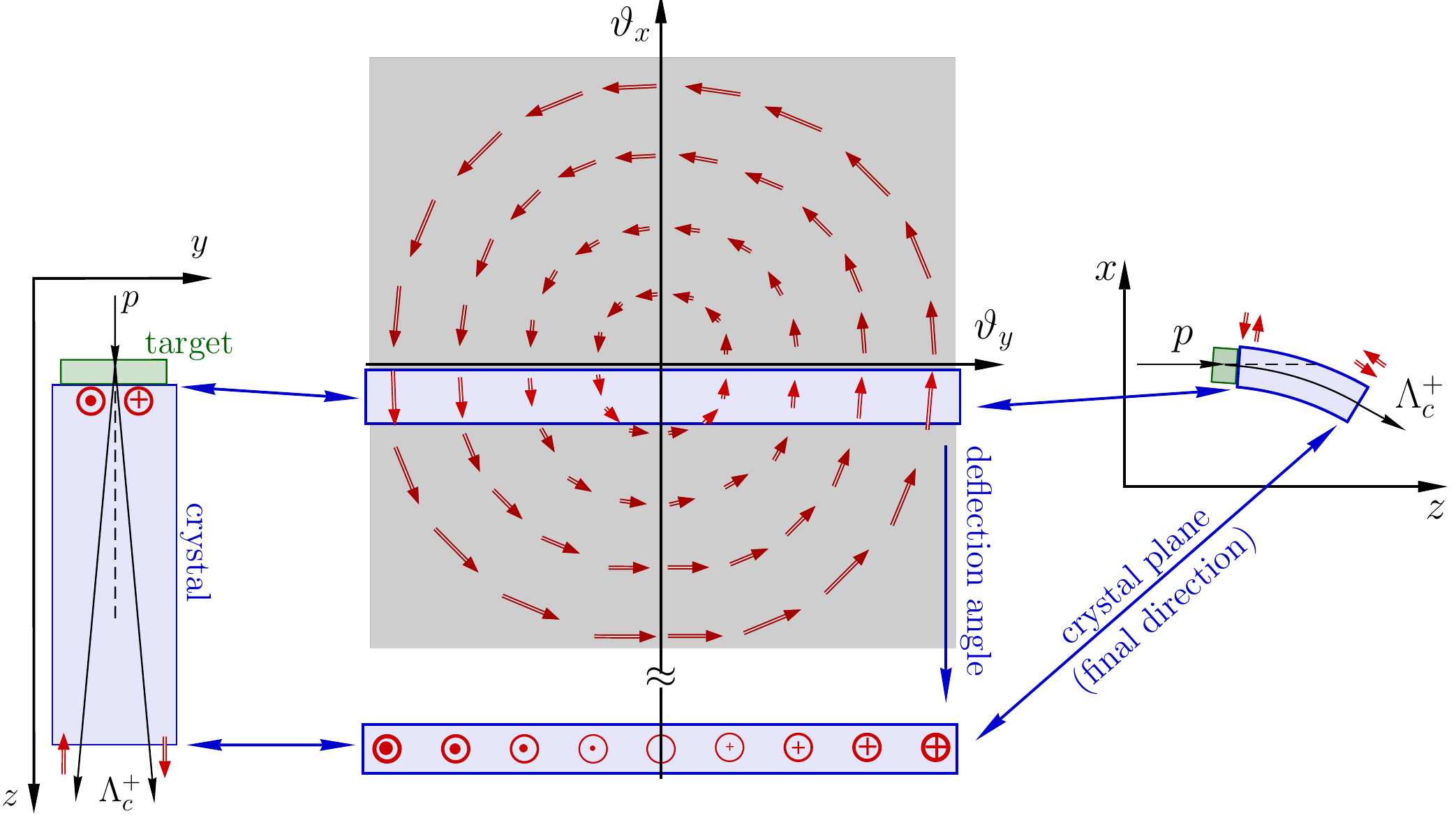}
\end{center}
\caption
{
Selection of $\Lambda_c$ initial polarisation by the crystal, and spin precession in a bent crystal (for the case $\phi=\pi/2$).
(Middle) the distribution of the $\Lambda_c$ polarisation in the phase space $\vartheta_x\vartheta_y$, here $\vartheta=\sqrt{\vartheta_x^2+\vartheta_y^2}$ is the angle between the proton and the $\Lambda_c$ momenta.
Red arrows show the $\Lambda_c$ polarisation.
The blue rectangular areas close to $\vartheta_y$ axis and at the bottom of the plot show the phase spaces of initially captured and deflected $\Lambda_c$ baryons, respectively.
The layout of the target-crystal setup (left) in the $yz$ plane and (right) in the $zx$ plane.
}
\label{fig:polarisation}
\end{figure}

We propose to place a crystal immediately after the target, to deflect as many $\Lambda_c$ baryons as possible before they decay (see Fig.~\ref{fig:polarisation}, left and right).
Note that the crystal selects by channeling only a small fraction of produced baryons, that have a small angle with respect to the crystallographic plane,
$\vartheta_x\in(\vartheta_{\rm crys}-\vartheta_{\rm acc}, \vartheta_{\rm crys}+\vartheta_{\rm acc})$.
Here $\vartheta_{\rm acc}$ is the acceptance angle to channeling~\cite{Lindhard:1965, Bezshyyko:2017var}, the
axis $Oz$ is chosen in the direction of impinging protons,
the axis $Oy$ lies in the crystal plane,
the initial direction of crystal plane normal is shifted from $Ox$ by a small angle $\vartheta_{\rm crys}$ around $Oy$ axis, and the crystal is bended around the $Oy$ direction (see Fig.~\ref{fig:polarisation}).

For the MDM measurement the optimal orientation is when the crystallographic plane is aligned with the impinging proton beam ($\vartheta_{\rm crys}=0$), as in this case the $x$-component of polarisation, ${\rm i.e.}$ orthogonal to spin precession axis, is maximal.
Note that with this orientation $\Lambda_c$ initial polarisation is almost parallel to the $\vec{n}_x$ axis, with two fractions that are positively or negatively polarised and can be separated experimentally by reconstructing $\vartheta_y$.
This feature was used to cancel the systematic uncertainty connected with the acceptance of the apparatus in Fermilab experiment E761~\cite{Chen:1992ai}.

In the current study we considered two setup configurations proposed in~\cite{Mirarchi:2019ft}, that are (IR3) at momentum cleaning area of LHC and (IR8) in front of interaction point at the LHCb detector.
In both cases the target and the crystal materials are tungsten and silicon, respectively, and the target length is $5\,$mm.
The other parameters of the setup are presented in Table~\ref{tab:targ-crys}.

As the $\Lambda_c$ production is a rare event, it is important to avoid channeling of initial protons, ${\rm i.e.}$ ones that pass through the target with negligible interaction.
In the case of experiment at the extraction line this can be done by slight ($\vartheta_{\rm crys}=100$--$200\, \mu$rad) misalignment of the crystal.
As we show latter, this misalignment would have no effect on the measurement efficiency (see Fig.~\ref{fig:f(Q)}).
For the circulating machine the risk of accidental deflection of the initial protons to the beam pipe is not tolerable, so the bending radius of the crystal is chosen in order to avoid channeling at the top energy~\cite{Mirarchi:2019ft}.

\begin{figure}[t]
\begin{center}
\includegraphics[width=0.99\textwidth]{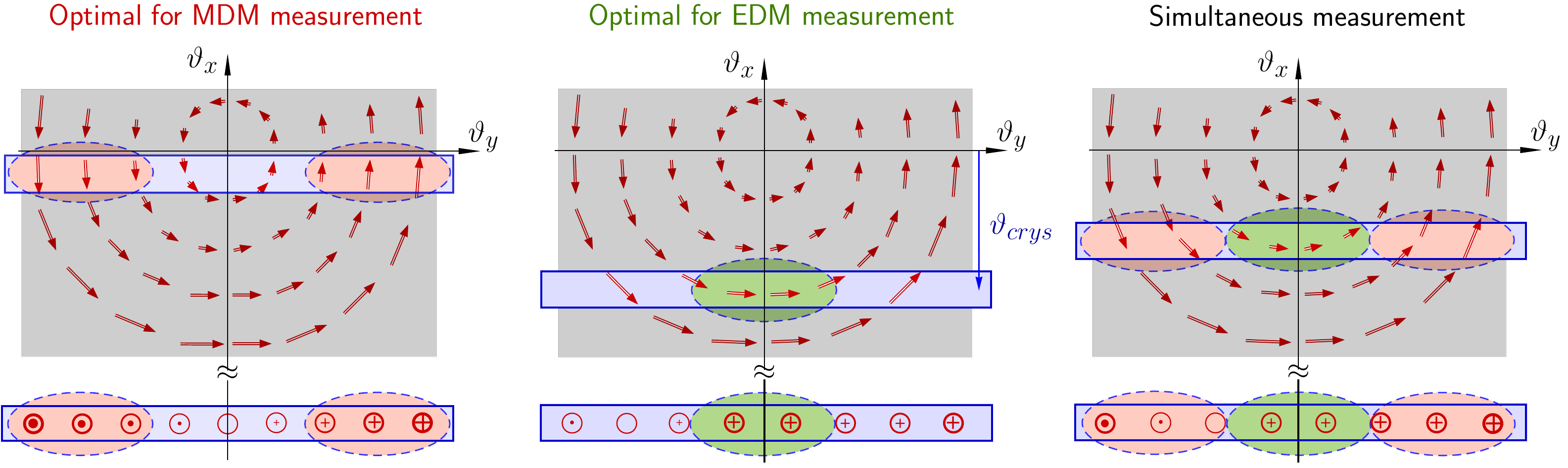}
\end{center}
\caption
{Same as in Fig.~\ref{fig:polarisation}.
The areas highlighted with red and green show the phase spaces of $\Lambda_c$ baryons with spin precession in crystal caused mainly by MDM and EDM, respectively.
}
\label{fig:pol_EDM}
\end{figure}

The $\Lambda_c$ electric dipole moment (EDM) can be obtained by measuring the spin precession caused by interaction of particle EDM with the electric field of crystal planes~\cite{Botella:2016ksl}.
For this measurement the initial polarisation should have considerable component perpendicular to crystal electric field~\cite{Baryshevsky:2019vou}, ${\rm i.e.}$ $y$-component.
Note that to achieve this condition the crystal should be rotated by a small angle $\vartheta_{\rm crys}$ around $Oy$-axis, but not by $90^\circ$ around $Oz$-axis,
as shown in Fig.~\ref{fig:pol_EDM}.
Here we present a simplified scheme just to demonstrate the main direction of spin precession caused by interaction of $\Lambda_c$ MDM and EDM with electric field of bent crystal.
One can see that by orienting the crystal with respect to the initial proton beam we can select deflected $\Lambda_c$ baryons with a certain (parallel or perpendicular to crystallographic plane) initial polarisation. 
Thus there are three possible initial crystal orientations optimised for MDM, EDM and simultaneous measurement, presented in Fig.~\ref{fig:pol_EDM} left, centre and right, respectively.
The phase space in the blue rectangles at the bottom represent the deflected $\Lambda_c$ baryons.
To simplify the picture we suppose that the precession angle is $\pi/2$ and the final polarisation has only z-component.
We consider this in more details in the following section.
In \cite{Bezshyyko:2017var} it was shown that it should be easy to separate experimentally deflected $\Lambda_c$ baryons events since the deflection angle is greater than the typical production angle of the deflected part of $\Lambda_c$ baryons. Note that it is also very important to reconstruct $\vartheta_y$ especially for MDM measurement.

\begin{figure}[t]
\begin{center}
\includegraphics[width=.99\textwidth]{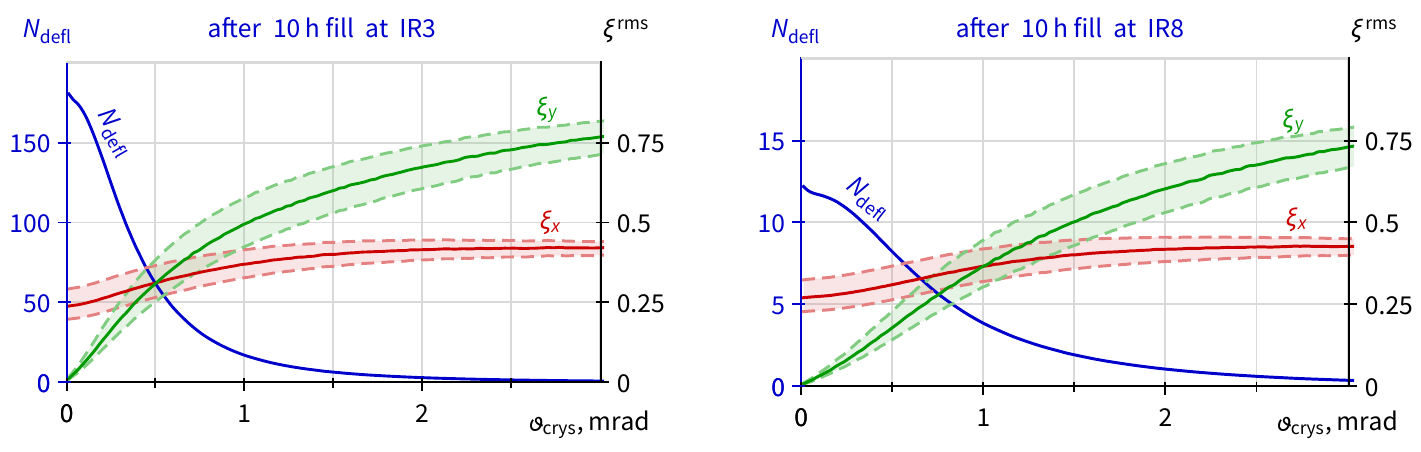}
\end{center}
\caption{
Root mean square of initial polarisation of $\Lambda_c$ deflected by the crystal ($\xi_x$ and $\xi_y$ --- projections on axes $Ox$ and $Oy$, respectively) and $N_{\rm defl}$ --- the integral number of deflected $\Lambda_c$ baryons after 10 hour LHC fill as functions of initial crystal orientation $\vartheta_{\rm crys}$.
(Left) and (right) for configurations at IR3 and IR8, respectively.
}
\label{fig:P(Q)}
\end{figure}

To calculate the average initial polarisation of deflected $\Lambda_c$ baryons and to verify the optimal crystal orientation, we performed computer simulations of $\Lambda_c$ propagation though a crystal using the approach described in~\cite{Bezshyyko:2017var, FominThesis}.
The results are presented in Fig.~\ref{fig:P(Q)} and in Table~\ref{tab:targ-crys}.

\subsection{Final polarisation}

Due to the MDM, the spin precession takes place in the $xz$ plane.
We first choose the polarisation axis to be perpendicular to the production plane, ${\rm i.e.}$
{$\vec{n}_x\equiv \frac{\vec{p}_{\rm beam}\times \vec{p}_{\Lambda_c}}{|\vec{p}_{\rm beam}\times \vec{p}_{\Lambda_c}|}$}.
In this case, supposing $\vartheta_{\rm crys}=0$, we can write the polarisation of the $\Lambda_c$ before going through the crystal by the absolute value of the polarisation:
\[\xi\arrowvert_{\vec{n}_x}=\pm|\xi_x|=\pm|\xi|\]
Two signs correspond to two fractions of $\Lambda_c$ baryons of positive and negative initial polarisations.
After passing through the crystal, the $\Lambda_c$ spin precesses in the plane perpendicular to the effective magnetic field $\vec{B}$.
As a result, the polarisation of $\Lambda_c$ after the crystal  is modified as: 
\begin{equation}
\xi|_{\vec{n}_x}=\pm|\xi_x|\cos\phi 
\label{eq:Xi_x}
\end{equation}

If we choose another polarisation axis (let us call it $\vec{n}_z$), that is perpendicular to $\vec{n}_x$ and to the effective magnetic field $\vec{B}$ the polarisation after crystal is modified as
\begin{equation}
\xi|_{\vec{n}_z}=\pm|\xi_x| \sin\phi.
\label{eq:Xi_z}
\end{equation}

If we rotate the crystal by a small angle $\vartheta_{\rm crys}$ around $Oy$-axis, the $y$-component of initial polarisation $\xi_y$ will appear (see Fig.~\ref{fig:P(Q)}).
This would provide a better efficiency for the EDM measurement with respect to the setup considered in~\cite{Botella:2016ksl}. 
Indeed, using spin precession equation \cite{Bargmann:1959gz,Metodiev:2015gda} one can show that, if the particle possesses an EDM, its interaction with the average electric field of bent crystal will cause the spin precession around $Ox$ axis.
Assuming $\gamma \gg 1\ {\rm and}\ g\approx2$ 
\begin{eqnarray}
\xi|_{\vec{n}_y}&\approx&	\xi_y \cos \phi^\prime,	\\	
\xi|_{\vec{n}_z}&\approx&	\xi_y \sin  \phi^\prime,
\qquad\phi^\prime  \approx	  \frac{\,\omega\,\gamma\,f\,}{2},
\label{eq:precessionEDM}
\end{eqnarray}
where $\xi_y$ is the $y$-component of initial polarisation, $\phi^\prime$ is the precession angle around $Ox$ axis, and $f$ is a dimensionless EDM.

In case $g\neq 2$, there might be a significant spin rotation around $Oy$ which would be important to take into account.
By taking a theoretical prediction of $g$-factor $g=1.92$ and other parameters from the Table~\ref{tab:targ-crys} and pluging them into equation~\eqref{eq:precession}, we get the values for spin rotation angle around $Oy$ axis: $8.5^\circ$ and $16.3^\circ$ for configurations at IR3 and IR8, respectively.
This can be translated to a 5--9$\,\%$ correction to the equations~\eqref{eq:Xi_x}--\eqref{eq:precessionEDM} due to interaction of precessions caused by MDM and EDM.
Note that this effect can be mitigated by comparing two fractions of $\Lambda_c$ with positive and negative $\vartheta_y$.

\subsection{MDM and EDM measurement accuracy}

Analysing the angular distribution~\eqref{eq:AD} and considering~\eqref{eq:precession},~\eqref{eq:Xi_x}--\eqref{eq:precessionEDM} one can obtain the expressions for the uncertainty to the $\Lambda_c$ baryon $g$-factor and dimensionless EDM: 
\begin{equation}
 \Delta g = \sqrt{\frac
 		{12}
		{\ \alpha_j^2 \, Br_j \, \eta^{\rm det}_j
		\ \ N_{\rm defl}
		\ \eta_{\rm MDM}
		}},
		\qquad
		\eta_{\rm MDM}=
				\,\langle
				\,\xi_{x}^2
				\,\gamma^2
				\,\rangle
				\,\omega^2,
\label{eq:dg-fixE}
\end{equation}
\begin{equation}
 \Delta f = \sqrt{\frac
 		{12}
		{\ \alpha_j^2 \, Br_j \, \eta^{\rm det}_j
		\ \ N_{\rm defl}
		\ \eta_{\rm EDM}
		}},
		\qquad
		\eta_{\rm EDM}=
				\,\langle
				\,\xi_{y}^2
				\,\gamma^2
				\,\rangle
				\,\omega^2,
\label{eq:dg-fixE}
\end{equation}
where $\alpha_j$, $Br_j$ and $\eta_j^{\rm det}$ are the weak decay parameter, the branching ratio and the detector efficiency for $j$ decay channel,
$\eta_{\rm MDM}$ and $\eta_{\rm EDM}$ are the efficiencies of crystal-target setup for MDM and EDM measurement, respectively.
First three terms depend only on the $\Lambda_c$ decay channel and on the detector efficiency.
Last two terms are defined mainly by channeling efficiency and properties of the accelerator.
We call the product of these two parameters the precession efficiency. Its maximum corresponds to the optimal crystal configuration for the measurement.

\begin{figure}[b]
\begin{center}
\includegraphics[width=.95\textwidth]{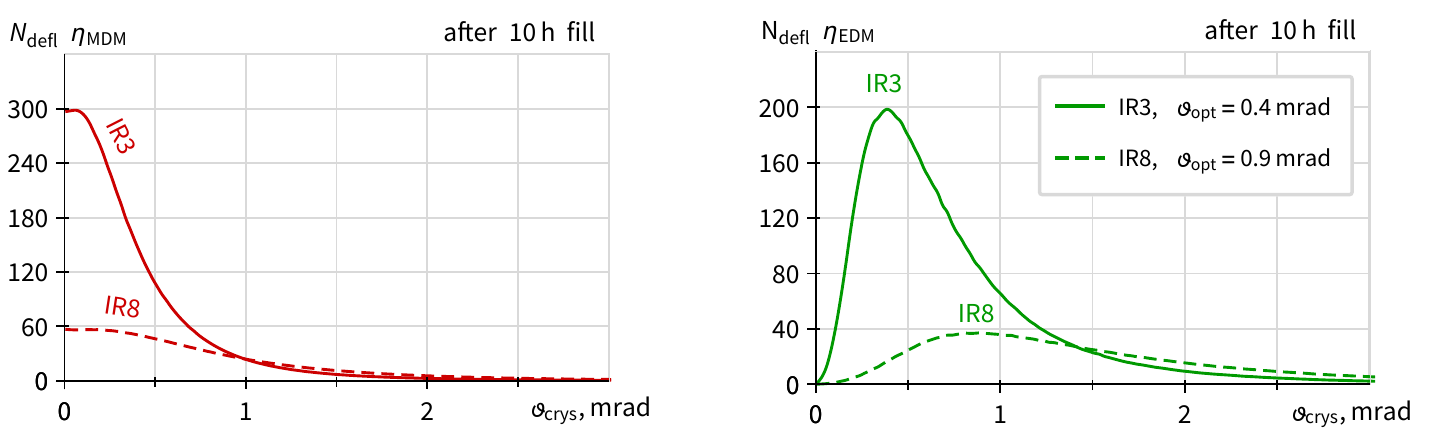}
\end{center}
\caption
{
Precession efficiency (left) of MDM and (right) of EDM measurements as a function of crystal orientation $\vartheta_{\rm crys}$. The same configurations as in~Fig.~\ref{fig:P(Q)}.
}
\label{fig:f(Q)}
\end{figure}

The results of computer simulations show that for MDM measurement the optimal initial orientation of the crystal is $|\vartheta_{\rm crys}|\leq0.15\,$mrad and $|\vartheta_{\rm crys}|\leq0.3\,$mrad for configurations at IR3 and IR8, respectively; and for EDM measurement: $\vartheta_{\rm crys}\approx0.4\,$mrad and $\vartheta_{\rm crys}\approx0.9\,$mrad (see Fig.~\ref{fig:f(Q)}).
The difference between last two angles is due to a softer spectra of deflected $\Lambda_c$ baryons at IR8. The baryons with the same transverse momentum $p_t$ (same polarisation $\xi$) but with smaller energy would have a greater production angle $\vartheta$.

The precession efficiencies of MDM and EDM measurements at IR3 are $\sim5.3$ times better.
This is because the setup at IR8 is limited by the properties of LHCb detector, whereas for IR3 the optimal parameters of the detector (acceptance angle, energy range, etc.) were obtained in order to maximise the double crystal efficiency.
The obvious downside of the IR3 configuration is that it requires building a new detector, but on the other hand, as it would be dedicated to this measurement, the detecting efficiency of the particular events could be much better than at LHCb.
The values of precession efficiencies and properties of deflected $\Lambda_c$ baryons are listed in the Table~\ref{tab:targ-crys}.

\begin{table}[t]
\caption
{
Parameters of target-crystal setup and properties of deflected $\Lambda_c$ baryons at IR3 and IR8.
}

\begin{center}
\begin{tabular}{|c| l |c|c|}
  \hline
&&~~IR3~~&~~IR8~~\\
\hline\multirow{4}{*}{\begin{tabular}{c}{~Target-crystal setup~}{\cite{Mirarchi:2019ft}}\end{tabular}}
&~Number of protons on target per $10\,$h fill&$~~3\times10^{10}$~~&~~$4.3\times10^{10}$~~\\
&~Target lenght, cm                         & 0.5 & 0.5 \\
&~Crystal length, cm                         &   7 & 7.5  \\
&~Crystal bending radius, m             & 14 & 5.4   \\
&~Deflection angle $\omega$, mrad &   5 & 14     \\
\hline\multirow{6}{*}{\begin{tabular}{c}Deflected $\Lambda_c$ baryons\\(optimised for MDM)\end{tabular}}
&~Initial crystal orientation                                 $\vartheta_{\rm crys}$, mrad  & 0.1            & 0.1          \\
&~Number of deflected $\Lambda_c$ per $10\,$h fill      $N_{\rm defl}$          & 180           & 12             \\
&~Average $\Lambda_c$ Lorentz factor                          $\overline\gamma$  & 1140          & 600             \\
&~Expected spin rotation angle  (for $\overline g=1.92$) $\phi$                      &$-8.5^\circ$&$-16.3^\circ$ \\
&~Average $\Lambda_c$ polarisation ($x$-component) $\xi_x^{\rm rms}$     & 0.24(5)       & 0.27(5)          \\
&~Weighted average polarisation 
                    $\sqrt{\langle\xi_x^2 \gamma^2\rangle/\langle\gamma^2\rangle}$&0.22(5)       &0.26(5)              \\
&~Precession efficiency (per $10\,$h fill),   $N_{\rm defl}\ \eta_{\rm MDM}$~~& 300            & 57                      \\
\hline\multirow{5}{*}{\begin{tabular}{c}Deflected $\Lambda_c$ baryons\\(optimised for EDM)\end{tabular}}
&~Initial crystal orientation                                 $\vartheta_{\rm crys}$, mrad & 0.4      & 0.9  \\
&~Number of deflected $\Lambda_c$ per $10\,$h fill,     $N_{\rm defl}$         & 75       & 5       \\
&~Average $\Lambda_c$ Lorentz factor                          $\overline\gamma$ & 910     & 570     \\
&~Average $\Lambda_c$ polarisation ($y$-component) $\xi_y^{\rm rms}$    & 0.25(5)& 0.34(5) \\
&~Weighted average polarisation
                  $\sqrt{\langle\xi_y^2 \gamma^2\rangle/\langle\gamma^2\rangle}$&0.34(5)&0.41(5)      \\
&~Precession efficiency (per $10\,$h fill),  $N_{\rm defl}\ \eta_{\rm EDM}$~~& 200      & 37            \\
\hline
\end{tabular}
\end{center}
\label{tab:targ-crys}
\end{table}

Note that the  $\Lambda_c$ baryons with different directions of polarisation can be separated by reconstructing $\vartheta_y$.
${\rm E.g.}$ at $\vartheta_y=0$ and $\vartheta_x\neq0$ polarisation has only $y$-component (see Fig.~\ref{fig:polarisation}, middle), which makes it the optimal region for EDM measurement. 
At the same time, with the same crystal orientation $\vartheta_{\rm crys}$ but with $\vartheta_y\geq\vartheta_x$ the polarisation has also a considerable $x$-component which is essential for MDM measurement (see Fig.~\ref{fig:pol_EDM}, left).
Thus the measurement of MDM and EDM can be done at the same time with a small $\sim20\,\%$ drop of efficiency, with respect to the optimal one for each of them (see Fig.~\ref{fig:f(Q)}).
In this case the crystal should be oriented at $\vartheta_{\rm crys}=0.25\,$mrad and $\vartheta_{\rm crys}=0.6\,$mrad for measurement at IR3 and IR8, respectively.

\begin{figure}[t]
\begin{center}
\includegraphics[width=.47\textwidth]{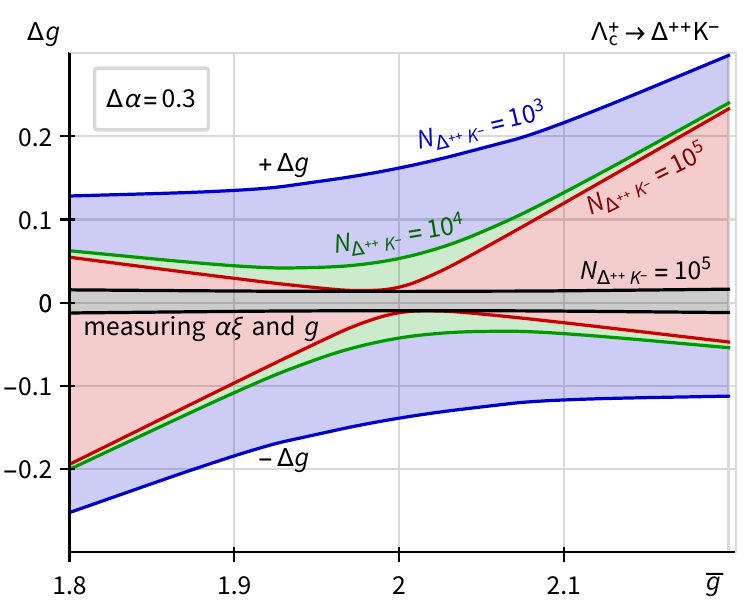}~~~~~~
\includegraphics[width=.47\textwidth]{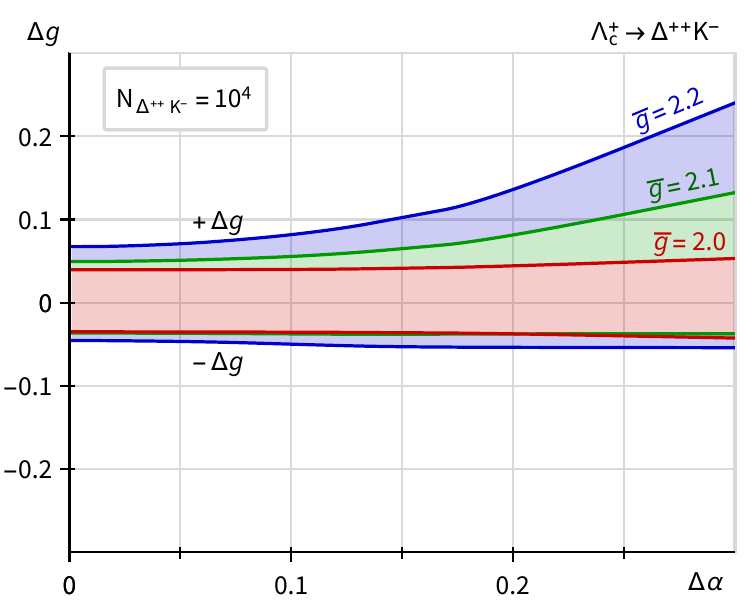}
\end{center}
\caption{Absolute statistical error of $g$-factor
(left) as a function of its expected value $\overline g$, and
(right) as a function of error on alpha parameter $\Delta\alpha$.
(Left) for different numbers of registered events and
(right) for various expected values of $g$-factor $\overline g$, stated in the plot.
Calculation results for IR3 configuration, considering $\Lambda_c^+\to \Delta^{++}K^-$ decay channel, assuming $\alpha=0.67$.}
\label{fig:dg_g}
\end{figure}

Another very important parameter for reconstruction of final polarisation is a weak decay parameter.
Below, we consider only the MDM measurement, but the approach can be extended also to the EDM measurement.
In Fig.~\ref{fig:dg_g}, we demonstrate the sensitivity of $g$-factor precision to a weak decay parameter uncertainty $\Delta\alpha$.
One can see that the poor knowledge of $\Delta \alpha= 0.3$ essentially limits the precision of the MDM as it adds a significant systematical uncertainty.

On the other hand, this problem can be solved by measuring the $\alpha\,\xi_x^{\rm rms}$ factor and the MDM at the same time.
If we neglect EDM, the spin precession modifies the polarisation projection on both axes, $\vec{n}_z$ and $\vec{n}_x$, which can be measured independently~\cite{Chen:1992ai},
\begin{equation}
\frac{1}{N}\frac{dN}{d\cos \theta_x}=\frac{1}{2}(1+\alpha\,\xi_x^{\rm rms} \cos\phi \cos \theta_x), 
\quad 
\frac{1}{N}\frac{dN}{d\cos \theta_z}=\frac{1}{2}(1+\alpha\,\xi_x^{\rm rms} \sin\phi \cos \theta_z),
\end{equation}
where	$\cos \theta_x = \vec{n}_x\cdot \vec{n}_{\rm baryon}$
and		$\cos \theta_z = \vec{n}_z\cdot \vec{n}_{\rm baryon}$.
Thus, we can have two observables: two angular coefficients
$b_x \equiv \alpha\,\xi_x^{\rm rms} \cos\phi $ and
$b_z \equiv \alpha {\xi_x^{\rm rms}} \sin\phi $ that can provide both
$\alpha \,\xi_x^{\rm rms}$ and the $\phi$ angle via 
\begin{equation}
b_x^2+b_z^2	=	\alpha^2\,\langle\xi^2\rangle,\qquad
\frac{b_z}{b_x}	=	\tan\phi.
\end{equation}
The uncertainty of $g$-factor in this case is
\begin{equation}
 \Delta g = 
  \frac{1}
         {\overline \alpha_j 
         \,\xi_x^{\rm rms}
           \,\gamma
             \,\omega~}
 \ \sqrt{\frac{12}
                  {~N_j~} }
                  \left(1+\sqrt{2}\,\frac{|\overline g - 2|}{2}\omega\gamma\right),\qquad 
                  N_j=Br_j\,\eta_j^{\rm det}\,N_{\rm defl}.
\label{eq:dg-fixE}
\end{equation}
where $N_j$ is the number of reconstructed events.
The expression in parentheses represents the increase of error on $g$-factor due to simultaneous measurement of $\alpha\,\xi_x^{\rm rms}$. For $\overline g=1.92$ this factor is about 1.32 and 1.48 for IR3 and IR8 configurations, respectively.

\begin{figure}[t]
\begin{center}
\includegraphics[width=.99\textwidth]{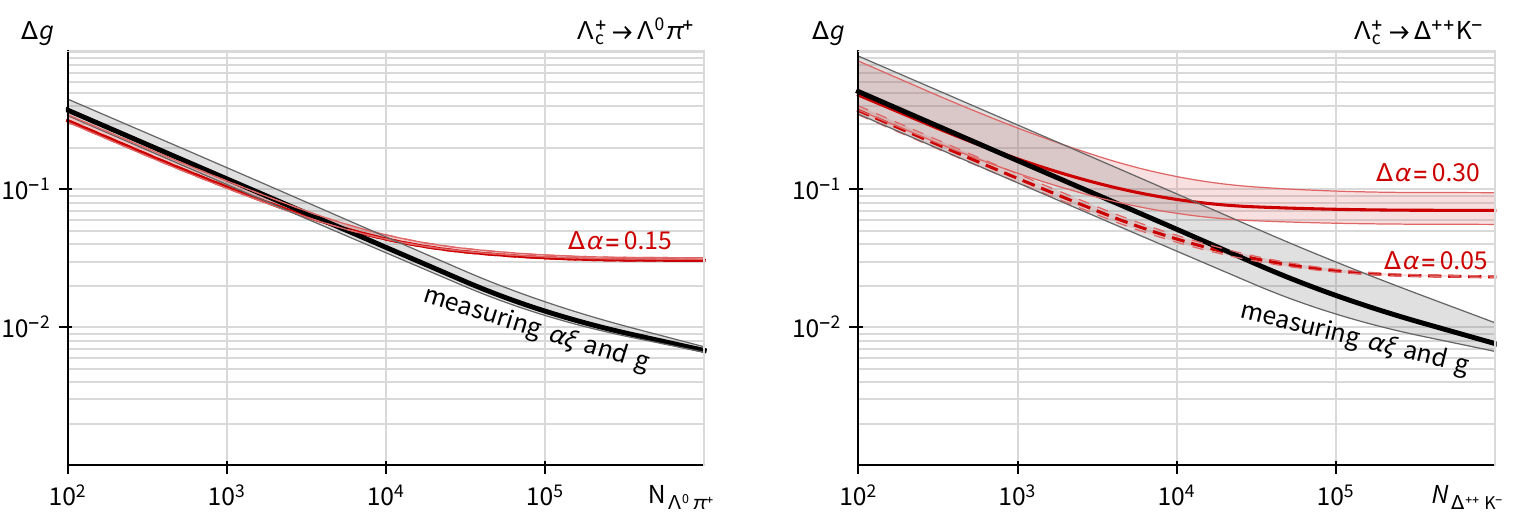}
\end{center}
\caption{Absolute statistical error of $g$-factor as a function of reconstructed events number $N_j$
(left) for $\Lambda_c^+\to \Lambda^0\pi^+$ and
(right) for $\Lambda_c^+\to \Delta^{++}K^-$ decay channels.
Red curves labeled with $\Delta\alpha$ value are obtained using the pre-measured values of $\alpha\,\xi_x^{\rm rms}$ factor,
black line corresponds to the case when we measure $\alpha\xi$ and $g$-factor simultaneously at current experiment.
Calculation results for IR3 configuration. Margins represents the current uncertainly on a weak decay parameter value.
}
\label{fig:dg_N}
\end{figure}

The figure~\ref{fig:dg_N} presents the evolution of uncertainty to $g$-factor with number of reconstructed events, ${\rm i.e.}$ when $\Lambda_c$ baryon is deflected by a full bending angle of the crystal and then decays by a certain channel stated in the figure.
Here we compare two cases: when $g$-factor is measured while the value $\alpha\,\xi_x^{\rm rms}$ is taken from the another experiment (red curves labeled with $\Delta\alpha$ values) and when $g$-factor and $\alpha\,\xi_x^{\rm rms}$ are measured simultaneously (black curves).
In the latter case we assume the expected value of weak decay parameter to be 
$\overline\alpha=0.91$ for $\Lambda_c^+\to \Lambda^0\pi^+$ decay (left) and 
$\overline\alpha=0.67$ for $\Lambda_c^+\to \Delta^{++}K^-$ decay, 
$\overline g=2.1$, and other parameters of IR3 configuration (see Table~\ref{tab:targ-crys}).

One can see that using the external value of $\alpha\,\xi_x^{\rm rms}$ improves the precision to $g$-factor while $\Delta g\gtrsim0.1$, or at low statistics ($10^3$--$10^4$ events), and after collecting more data the systematical error from $\Delta\alpha$ becomes dominant and it is more efficient to measure two factors at the same time.
The other potential source of systematic error is expected from $\Lambda_c$ energy reconstruction.
Our calculations show that this impact is quite small, ${\rm e.g.}$ even for the energy error as big as $\Delta\varepsilon=100\,$GeV we start to see the effect on $\Delta g$ only after $10^5$ events (see the bend of black curves in Fig.~\ref{fig:dg_N}).

In any case it is very important to know $\alpha\,\xi_x^{\rm rms}$ as ithe current precisions for $\Lambda_c\to p\  K^*$ and $\Lambda_c^+\to \Delta^{++}K^-$ channels give almost one order of magnitude uncertainty on data taking time needed to reach to a certain $\Delta g$. 
The factor $\alpha\,\xi_x^{\rm rms}$ could be pre-measured if we can have exactly the same setup for the $\Lambda_c$ production: using the fixed-target data sample collected at the LHCb experiment with the SMOG system \cite{Aaij:2014ida} might be an interesting possibility.
On the other hand, to reach to a higher precision, we may need to obtain this factor from other experiments, such as LHCb which has a much higher statistics.
As the $\alpha$ value is the same in any environment, this can be measured precisely by LHCb.
Per contra, the $\xi_x^{\rm rms}$ value depends on $p_{\rm t}$ and $\Lambda_c$ energy $\varepsilon$, so we need a theoretical extrapolation, which leads to some uncertainties.
In any case, if we use the LHCb data, the $\alpha$ parameters should be reconstructed separately.

We discuss how we can achieve that in the next section, considering  two decay processes, $\Lambda_c\to p \pi \pi$ and $\Lambda_c \to p K \pi$.
The first decay is intermediated by the $\Lambda_c\to \Lambda \pi$ whose $\alpha$ value has been measured as $\alpha= 0.91 \pm 0.15 $.
The second decay is more complex since there are three intermediate channels, $\Lambda_c\to K^*(890) p,\, \Delta^{++} K,\, \Lambda(1520)\pi\to p K \pi$, which introduce three different  $\theta$ angles (c.f. $\theta$ is defined by the direction of these intermediate baryons).
Despite of this complexity, the second decay may be able to determine MDM more precisely since it has a larger branching ratio comparing to the first one, which occurs via successive weak decays.
Another drawback of the first decay is the presence of relatively long-living $\Lambda^0$ baryon in the intermediate state, that significantly reduces the detecting efficiency at LHCb detector (by about $47$ times, according to~\cite{Bagli:2017foe}).
At IR3 this problem could be partially solved by building a longer detector, but as the average energy of deflected $\Lambda_c$ baryons is twice greater with respect to IR8, we do not expect the gain of more than 5 times with respect to IR8.

\begin{table}[t]
\caption
{
Properties of different decay channels of $\Lambda_c$ baryons.
}

\begin{center}
\begin{tabular}{| l |c|c|c|c|c|}
  \hline
~Decay channel &~ Branching ratio ~&~ Weak decay param. ~&\multicolumn{2}{c|}{Detector efficiency $\eta_j^{\rm det}$}&~ Wieght, ~\\
&$Br_j$, \% \cite{Tanabashi:2018oca}&$\alpha_j$ (see section~\ref{sec:alphas})&~~~~~~IR3~~~~~~&~~~ IR8~\cite{Bagli:2017foe}~~&~$(\Delta g/ \Delta g_j)^2~$\\
\hline
~$\Lambda_c^+\to p\         K^*(892)               $& 1.96(27)    &      0.66(28)      &  0.2  &      0.2   & $\sim0.60$\\
~$\Lambda_c^+\to \Delta^{++}(1232)  \, K^-$~& 1.08(25)    &$-$0.67(30)~~~&  0.2  &      0.2    & $\sim0.35$ \\
~$\Lambda_c^+\to \Lambda\, (p\ \pi^-)\ \pi^+ $& 0.83(5)~    &     0.91(15)      &~0.02&~~~0.004& 0.01--0.05 \\
~$\Lambda_c^+\to \Lambda   (1520)   \ \pi^+ $& 2.2(5)~~~ &$-$0.11(60)~~~&  0.2  &      0.2     &  $0.02$\\
\hline
\end{tabular}
\end{center}
\label{tab:decay}
\end{table}

In Table~\ref{tab:decay} we list the properties of the most useful $\Lambda_c$ decay channels in terms of polarisation reconstruction together with their detection efficiencies.
Using these values in Eq.~\eqref{eq:dg-fixE} we obtain the weights of these channels at MDM reconstruction (see Table~\ref{tab:decay} last column).
One can see that about $95\,\%$ of information for MDM reconstruction comes from the first two channels: $\Lambda_c\to K^*(890) p$ and $\Lambda_c^+\to\Delta^{++} (1232)K^-$.

\begin{figure}[t]
\begin{center}
\includegraphics[width=.99\textwidth]{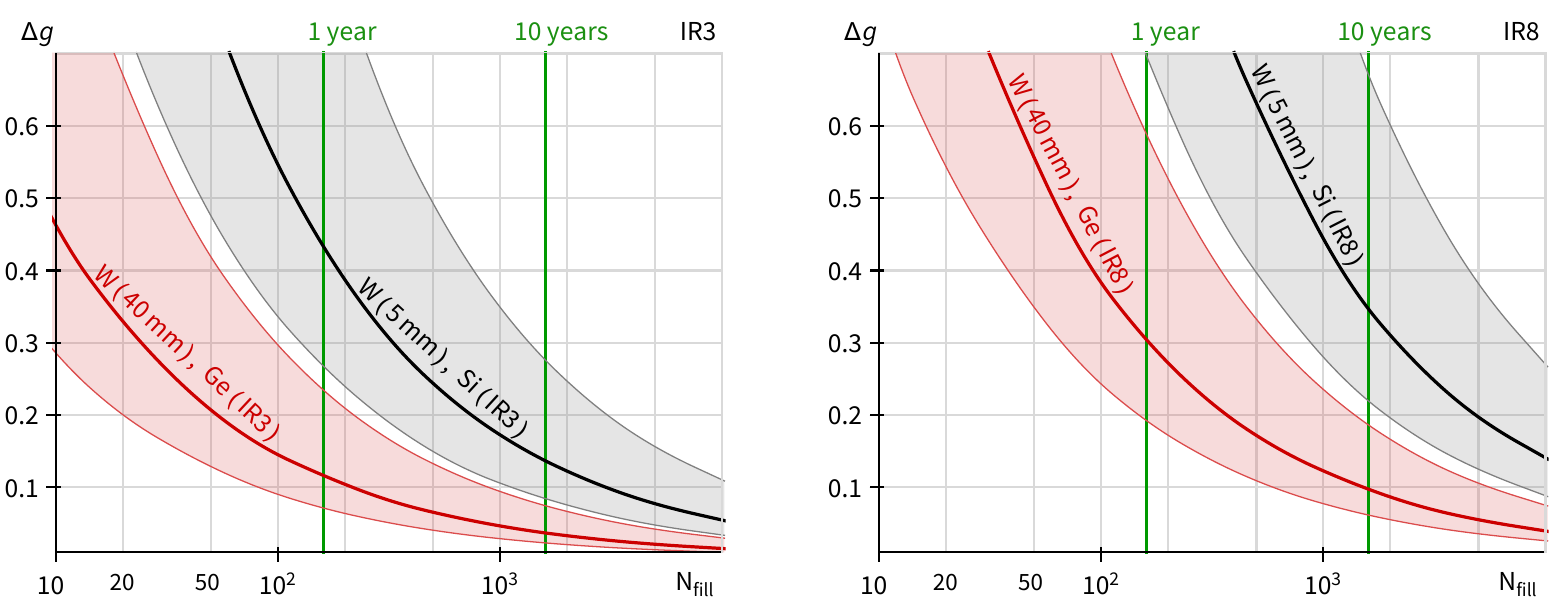}
\end{center}
\caption{Absolute statistical error of $g$-factor as a function of number of fills 
(left) for IR3 and
(right) for IR8 configurations.
Black curves for $5\,$mm tungsten target attached to silicon crystal,
red curves for $40\,$mm tungsten target attached to germanium crystal.
Margins represents the current uncertainly of the weak decay parameters and the initial polarisation.
}
\label{fig:dg_Nf}
\end{figure}

Finally, in Fig.~\ref{fig:dg_Nf} we present the absolute statistical error of the $g$-factor as a function of number of 10 hour LHC fills.
Due to a poor current knowledge of weak decay parameters and polarisation, the uncertainty of data taking time needed to reach the same $\Delta g$ is about one order of magnitude.
Two vertical green lines corresponds to 1 and 10 years of data taking based on LHC 2018 operation, for the consistency with \cite{Mirarchi:2019ft}.
The central values of $\Delta g$ for this time stamps are listed in the Table~\ref{tab:dg}.

\begin{table}[t]
\caption
{
Absolute statistical error of $g$-factor after 1 and 10 years of data taking for various configurations.
Data taking time needed to reach $\Delta g=0.1$ and $\Delta g =0.04$ (last two columns).
}

\begin{center}
\begin{tabular}{|c|c|c|c|c|c|c|}
\hline
\multicolumn{3}{|c|}{Configuration}&
\multicolumn{2}{c|}{~~~~~~~$\Delta g\,$ after~~~~~~~~}&
\multicolumn{2}{|c|}{~ Time (years) to reach ~}\\
\hline
~~Target length~~&~~~~~~Crystal~~~~~~&~~~~Place~~~~&
~~~~1 year ~~~~&~~ 10 years ~~&
~ $\Delta g=0.1$ ~&$\Delta g=0.04$\\
\hline\multirow{2}{*}{~~$  5\,$mm ~~}&\multirow{2}{*}{ silicon        }& IR8 & 1.10 & 0.35 &123&--\\&& IR3 & 0.43 & 0.14 & 19&120\\
\hline\multirow{2}{*}{~~$40\,$mm ~~}&\multirow{2}{*}{ silicon        }& IR8 & 0.49 & 0.16 &  25&160\\&& IR3 & 0.17 & 0.06 &   3&19\\
\hline\multirow{2}{*}{~~$40\,$mm ~~}&\multirow{2}{*}{ germanium}& IR8 & 0.31 & 0.10 &  10&62\\&& IR3 & 0.12 & 0.04 &1.5&8.5\\
\hline
\end{tabular}
\end{center}
\label{tab:dg}
\end{table}

Our calculations show that in order to reach the error on $g$-factor at a few percents, the target length should be enlarged at least to $40\,$mm and the silicon crystal should be replaced with germanium, like it was suggested in~\cite{Bezshyyko:2017var}.
The length and the bending radius of germanium crystal for IR3 were chosen $7\,$cm and $10\,$m to avoid channeling of impinging protons, and for IR8 ($5\,$m and $3.3\,$cm) were taken from~\cite{Bagli:2017foe}.
Going from $5\,$mm to $40\,$mm target and switching to germanium reduces the data taking time by factors 6 and 2.4, respectively.
Further enlargement of the target should not essentially increase the efficiency because of the decay of $\Lambda_c$ and the shower productions inside of the target.

Using a dedicated detector in IR3 would give an additional reduction of data taking time by a factor of about 7.5 with respect to measurement at LHCb detector.

With the optimal orientation for EDM measurement obtained in this paper the error on dimensionless EDM $\Delta f$ is about $22\,\%$ greater than $\Delta g$.
Thus the EDM of $\Lambda_c$ baryon could be measured with an error $\sim2.6\,\times\,10^{-16}\,e\,$cm
using $40\,$mm tungsten target and germanium crystal after 10 years of data taking at IR8 or less than 2 years at IR3.
Note that in~\cite{Bagli:2017foe} the estimation of error on EDM is two orders of magnitude lower, but there the expected number of protons on target is 1400 greater, initial polarisation is 2.3 times greater and $g$-factor value 
is 1.4 whereas we consider more conservative prediction $g=1.92$.
Considering all this, the method proposed in \cite{Bagli:2017foe} is \mbox{$\sim6.5$ times} less efficient by precision or requires $\sim40$ times longer data taking time to reach the same precision.




\section{Improving the precision on weak asymmetry parameters of charmed baryons at LHCb}
\label{sec:alphas}

Eq.~(\ref{eq:AD}) shows that in the decay $\Lambda_c \to B \, P$,  the $\Lambda_c$ polarisation $\xi$ 
can not be measured separately from the parameter $\alpha$.
This problem can be solved if there are more observables (than just $\cos\theta$ dependence), 
which provides  independent information allowing to fit both $\alpha$ and $\xi$. 
In the following we introduce two such examples. 
\bigskip

\subsection{The case of $\Lambda_c \to \Lambda \pi$ followed by ${\Lambda} \to p \pi$}
 
Let us start with computing the first decay chain  $\Lambda_c\to \Lambda \pi$.
The parity violating interaction is induced by a weak interaction in the form
\begin{equation}
M_{\lambda_{\Lambda_c}, \, \lambda_{{\Lambda}}}=\overline{u}_{\Lambda}(p_{\Lambda}, \lambda_{\Lambda}) (A-B\gamma_5)
u_{\Lambda_c}(p_{\Lambda_c}, \lambda_{\Lambda_c}),
\end{equation}
where $p_{\Lambda_c}$ ($p_{\Lambda}$) is  the 4-momentum, 
and the constants $A$ and $B$ represent parity conserving and violating contributions, respectively.
The helicity $\lambda_{\Lambda_c}$ ($\lambda_{\Lambda}$) is the projection of the baryon spin in its momentum direction.

We next consider the subsequent decay $\Lambda\to p \pi$. The transition amplitude can be written similarly to the $\Lambda_c$ decay:
\begin{equation}
M_{\lambda_{\Lambda}, \, \lambda_p}=\overline{u}_{p}(p_p, \lambda_p) (a-b\gamma_5)u_{\Lambda}(p_{\Lambda}, \lambda_{\Lambda}).
\end{equation}

\begin{figure}[h]
\begin{center}
\includegraphics[width=0.54 \textwidth]{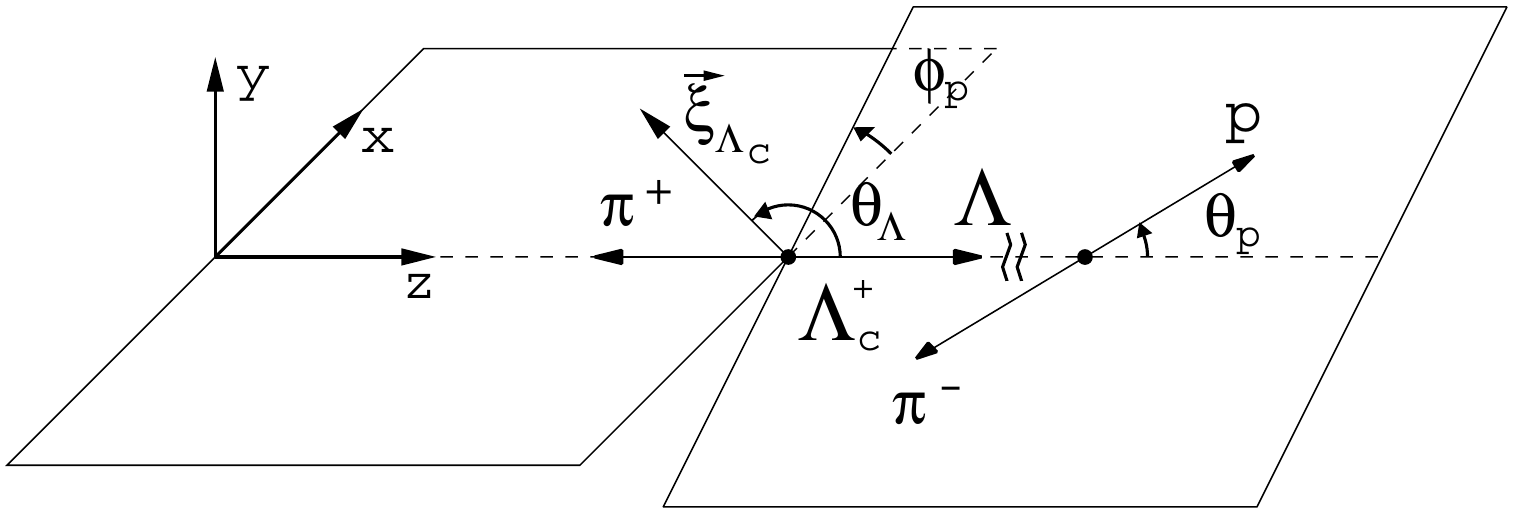}
\end{center}
\caption{Definition of angles in the polarised $\Lambda_c^+$ decay $\Lambda_c^+\to
\Lambda \,\pi^+ \to p \,\pi^- \, \pi^+$. }
\label{fig:polarised_Lambda_c}
\end{figure}

To describe the cascade decay $\Lambda_c^+\to
\Lambda \, \pi^+\to p \, \pi^- \, \pi^+$ of the polarised $\Lambda_c^+$ we choose the rest frame
of $\Lambda_c^+$.  In this frame the momentum of $\Lambda$ is directed along the $Oz$ axis,
and we assume that the polarisation vector $\vec{\xi}_{\Lambda_c}$ lies in the $xz$ plane with positive $x$-component
(see Fig.~(\ref{fig:polarised_Lambda_c})).
$\theta_\Lambda$ is the angle between $\vec{\xi}_{\Lambda_c}$ and $\Lambda$ momentum.
The polar angle $\theta_p$ is defined in the rest frame of $\Lambda$ baryon, and it is
the angle between the proton momentum and the $Oz$ axis. The azimuthal angle $\phi_p$ is the angle between
the decay plane $\Lambda \to p \pi^-$ and $xz$ plane.

The differential decay rate for $\Lambda_c^+\to \Lambda\pi^+\to p\pi^-\pi^+$ in this frame can be
written as
\begin{equation}
\frac{d\Gamma (\Lambda_c^+\to \Lambda\pi^+\to
p\pi^-\pi^+)}{d\cos\theta_\Lambda\,d\cos\theta_p\,d\phi_p}=\Gamma(\Lambda_c^+\to
\Lambda\pi^+) \, {\rm BR}(\Lambda\to
p\pi^-)\,W(\cos\theta_\Lambda\,,\cos\theta_p\,,\phi_p).\label{eq:00011}
\end{equation}
Here
\begin{eqnarray}
W(\cos\theta_\Lambda\,,\cos\theta_p\,,\phi_p)&=&\frac{1}{8\,\pi}\Bigl(1+\alpha_{\Lambda_c}\alpha_\Lambda\cos\theta_p+
\alpha_{\Lambda_c}\xi_{\Lambda_c}\cos\theta_\Lambda+
\alpha_\Lambda\xi_{\Lambda_c}\bigl(\cos\theta_\Lambda\cos\theta_p
\nonumber \\
&+&\gamma_{\Lambda_c}\sin\theta_\Lambda\sin\theta_p\cos\phi_p-\beta_{\Lambda_c}\sin\theta_\Lambda\sin\theta_p\sin
\phi_p \bigr)\Bigr) \label{eq:00012}
\end{eqnarray}
is the full angular distribution of this decay,
$\xi_{\Lambda_c}=|\vec{\xi}_{\Lambda_c}|$ and decay parameters for $\Lambda_c^+\to
\Lambda \,\pi^+ $ and  $\Lambda \to p \,\pi^- $ are defined as
\begin{eqnarray}
&&
\alpha_{\Lambda_c}=\frac{|A_+|^2-|A_-|^2}{|A_+|^2+|A_-|^2}=\frac{2\,{\rm
Re}\left(A_S^*A_P\right)}{|A_S|^2+|A_P|^2}, \qquad
\alpha_\Lambda=\frac{|a_+|^2-|a_-|^2}{|a_+|^2+|a_-|^2}=\frac{2\,{\rm
Re}\left(a_S^*a_P\right)}{|a_S|^2+|a_P|^2}, \nonumber \\
&& \beta_{\Lambda_c}=\frac{2\,{\rm
Im}\left(A_+A_-^*\right)}{|A_+|^2 + |A_-|^2}=\frac{2\,{\rm
Im}\left(A_S^*A_P\right)}{|A_S|^2+|A_P|^2}, \qquad
\gamma_{\Lambda_c}=\frac{2\,{\rm Re}\left(A_+A_-^*\right)}{|A_+|^2
+ |A_-|^2}=\frac{|A_S|^2-|A_P|^2}{|A_S|^2+|A_P|^2}.
\label{eq:0002}
\end{eqnarray}
Here
\begin{eqnarray}
&& A_+\equiv A_{\frac{1}{2}0}=A\,k_+^\prime+B\,k_-^\prime,\qquad A_-\equiv
A_{-\frac{1}{2}0}=A\,k_+^\prime-B\,k_-^\prime, \nonumber \\
&& A_S=A, \qquad A_P=\frac{k_-^\prime}{k_+^\prime}B, \nonumber \\
&& a_+\equiv a_{\frac{1}{2}0}=a\,k_++b\,k_-,\qquad a_-\equiv
a_{-\frac{1}{2}0}=a\,k_+-b\,k_-,\nonumber \\
&& a_S=a, \qquad a_P=\frac{k_-}{k_+}b, \label{eq:0003}
\end{eqnarray}
where $A_{\lambda \, 0 }$ ($a_{\lambda \, 0 }$) are helicity
amplitudes for the decay $\Lambda_c^+\to \Lambda \pi^+$
($\Lambda \to p \pi^-$), and $A_S$ ($a_S$) and $A_P$ ($a_P$) are the $S$- and $P$-wave amplitudes.
In addition
\begin{equation}\label{eq:0004}
k_\pm^\prime\equiv\sqrt{(m_{\Lambda_c}\pm m_\Lambda)^2-m_\pi^2},
\qquad k_\pm\equiv\sqrt{(m_\Lambda\pm m_p)^2-m_\pi^2}.
\end{equation}
The parameters $\alpha_{\Lambda_c}$, $\beta_{\Lambda_c}$, and $\gamma_{\Lambda_c}$ satisfy
\begin{equation}\label{eq:0005}
\alpha_{\Lambda_c}^2+\beta_{\Lambda_c}^2+\gamma_{\Lambda_c}^2=1.
\end{equation}
It is useful to introduce additional parameter $\Phi_{\Lambda_c}$
\begin{equation}\label{eq:0006}
\beta_{\Lambda_c}=\left(1-\alpha_{\Lambda_c}^2\right)^{1/2}\sin\Phi_{\Lambda_c}, \qquad
\gamma_{\Lambda_c}=\left(1-\alpha_{\Lambda_c}^2\right)^{1/2}\cos\Phi_{\Lambda_c}.
\end{equation}

Note that a formula similar to Eqs.~(\ref{eq:00011}), (\ref{eq:00012}) for differential decay rate has been written in Ref.~\cite{Korner:1991ap}, however the corresponding equation (20) in \cite{Korner:1991ap} has inaccuracies or misprints.

Having a sufficient number of events of the decay $\Lambda_c^+\to
\Lambda \,\pi^+ \to p \,\pi^- \, \pi^+$, and knowing the parameter
$\alpha_\Lambda = 0.642\pm0.013$~\cite{Tanabashi:2018oca}, one can use
Eq.~(\ref{eq:00012}) for estimation of the decay parameters $\alpha_{\Lambda_c}$, $\beta_{\Lambda_c}$,
$\gamma_{\Lambda_c}$ and $\Lambda_c^+$ polarisation
using, for example, the method of maximum likelihood.

From the general three-dimensional angular distribution, Eq.~(\ref{eq:00012}), one can obtain simpler
distributions. For example, by integrating Eq.~(\ref{eq:00012}) over the azimuthal angle, we get the
two-dimensional distribution
\begin{equation}
W_{\Lambda, \, p}(\cos\theta_\Lambda\,,\cos\theta_p) =
\frac{1}{4}\bigl(1+\alpha_{\Lambda_c}\alpha_\Lambda\cos\theta_p+
\alpha_{\Lambda_c}\xi_{\Lambda_c}\cos\theta_\Lambda+
\alpha_\Lambda \xi_{\Lambda_c} \cos\theta_\Lambda \cos\theta_p
 \bigr).
\label{eq:000121}
\end{equation}
This equation does not include parameters $\beta_{\Lambda_c}$ and $\gamma_{\Lambda_c}$ and its analysis
allows one to extract the asymmetry $\alpha_{\Lambda_c}$ and $\Lambda_c^+$ polarisation $\xi_{\Lambda_c}$.

If the number of events of $\Lambda_c^+\to \Lambda \,\pi^+ \to p \,\pi^- \, \pi^+$
is not sufficient, then for extraction of the unknown parameters in Eq.~(\ref{eq:00012})
one can use one-dimensional angular distributions which are obtained by integration of
(\ref{eq:00012}) over two angles.
In this way we obtain one-dimensional angular distributions in
$\cos\theta_\Lambda$, $\cos\theta_p$ and $\phi_p$,
\begin{equation}
W_\Lambda(\cos\theta_\Lambda)=\frac{1}{2}
\left(1+\alpha_{\Lambda_c}
\xi_{\Lambda_c}\cos\theta_\Lambda\right), \label{eq:0007}
\end{equation}
\begin{equation}
W_p(\cos\theta_p)=\frac{1}{2} \left(1+\alpha_{\Lambda_c}
\alpha_{\Lambda}\cos\theta_p\right), \label{eq:0008}
\end{equation}
\begin{eqnarray}
W_\phi(\phi_p)&=&\frac{1}{2\,\pi} \left(1+\frac{\; \pi^2}{4}
\alpha_{\Lambda}\xi_{\Lambda_c}\left(\gamma_{\Lambda_c}\cos\phi_p-\beta_{\Lambda_c}\sin\phi_p\right)\right)\nonumber
\\ &=&\frac{1}{2\,\pi} \left(1+\frac{\; \pi^2}{4}
\alpha_{\Lambda}\left(1-\alpha_{\Lambda_c}^2\right)^{1/2}\xi_{\Lambda_c}\cos\left(\phi_p+\Phi_{\Lambda_c}\right)\right).
\label{eq:0009}
\end{eqnarray}

The product $\alpha_{\Lambda_c} \xi_{\Lambda_c}$ can be found from
the distribution in Eq.~(\ref{eq:0007}) by measuring the forward-backward asymmetry of $\Lambda$ baryon in the rest
frame of $\Lambda_c^+$
\begin{equation}
A_{\rm
FB}^{(\Lambda)}=\frac{F_\Lambda-B_\Lambda}{F_\Lambda+B_\Lambda}
=\frac{1}{2}\alpha_{\Lambda_c} \xi_{\Lambda_c},
\label{eq:0010}
\end{equation}
where
\begin{equation}
F_{\Lambda}\equiv \int_0^1 W_{\Lambda} (\cos\theta_\Lambda) \, d\cos\theta_\Lambda, \qquad
B_\Lambda\equiv \int_{-1}^0 W_{\Lambda} (\cos\theta_\Lambda) \,
d\cos\theta_\Lambda.
\label{eq:0011}
\end{equation}
The study of the distribution in Eq.~(\ref{eq:0008}) in the rest frame of baryon $\Lambda$, with a
known value of $\alpha_{\Lambda}$, will allow one to measure the parameter $\alpha_{\Lambda_c}$.
Indeed,
\begin{equation}
A_{\rm FB}^{(\rm p)}=\frac{F_{\rm p}-B_{\rm p}}{F_{\rm p}+B_{\rm
p}} = \frac{1}{2}\alpha_{\Lambda_c} \alpha_{\Lambda},
\label{eq:0012}
\end{equation}
where
\begin{equation}
F_p\equiv \int_0^1 W_p (\cos\theta_p) \, d\cos\theta_p,  \qquad
B_p\equiv \int_{-1}^0 W_p (\cos\theta_p) \, d\cos\theta_p.
\label{eq:0013}
\end{equation}
As a result we can find the magnitude of $\Lambda_c^+$ polarisation
\begin{equation}
 \xi_{\Lambda_c}=\alpha_{\Lambda}\frac{A_{\rm
FB}^{(\Lambda)}}{A_{\rm FB}^{(\rm p)}}.
\label{eq:0014}
\end{equation}
In order to find the remaining parameters $\beta_{\Lambda_c}$ and $\gamma_{\Lambda_c}$
one can apply the angular distribution Eq.~(\ref{eq:0009}) in the azimuthal angle
$\phi_p$ in the rest frame of $\Lambda$. For example, by measuring the following asymmetries:
\begin{equation}
A_1 \equiv \Bigl(\int\limits_{0}^{\pi/2}d\,\phi_p-\int\limits_{\pi/2}^{3\pi/2}d\,\phi_p+\int\limits_{3\pi/2}^{2\pi}d\,\phi_p
 \Bigr)W_{\phi} (\phi_p) =
\frac{\pi}{2}\alpha_{\Lambda}\xi_{\Lambda_c}\gamma_{\Lambda_c},
\label{eq:0015}
\end{equation}
\begin{equation}
A_2 \equiv\Bigl(\int\limits_{0}^{\pi}d\,\phi_p-\int\limits_{\pi}^{2\pi}
d\,\phi_p
 \Bigr) W_{\phi} (\phi_p) =-\frac{\pi}{2}\alpha_{\Lambda}\xi_{\Lambda_c}\beta_{\Lambda_c}.
\label{eq:0016}
\end{equation}
Then it follows from Eqs.~(\ref{eq:0015}) and (\ref{eq:0016}) that
\begin{equation}
 \frac{A_2}{A_1}=-\frac{\beta_{\Lambda_c}}{\gamma_{\Lambda_c}}=-\tan\Phi_{\Lambda_c}. \label{eq:0017}
\end{equation}
Therefore by studying one-dimensional angular distributions, the information on the $\Lambda_c^+$
polarisation and parameters of the decay $\Lambda_c^+\to \Lambda \,\pi^+ $ can be obtained.

Another way of measuring the $\Lambda_c^+$ polarisation in the decay $\Lambda_c^+\to\Lambda\pi^+$
is based on relation
between polarisations of $\Lambda_c^+$ and $\Lambda$ (see, e.g., \cite{Tanabashi:2018oca}):
\begin{equation}
\vec{\xi}_{\Lambda}=\frac{(\alpha_{\Lambda_c}+\vec{n}_\Lambda \cdot
\vec{\xi}_{\Lambda_c}) \vec{n}_\Lambda +\beta_{\Lambda_c}[\vec{\xi}_{\Lambda_c}\times \vec{n}_\Lambda ]+
\gamma_{\Lambda_c}[\vec{n}_\Lambda \times[\vec{\xi}_{\Lambda_c} \times \vec{n}_\Lambda  ]]}{1+\alpha_{\Lambda_c}
\vec{n}_\Lambda \cdot
\vec{\xi}_{\Lambda_c}}, \label{eq:0018}
\end{equation}
where $\vec{n}_\Lambda$ is a unit vector in the direction of the $\Lambda$ hyperon,
$\vec{\xi}_{\Lambda_c}$ is polarisation of the $\Lambda_c^+$ in the $\Lambda_c^+$ rest frame.
$\vec{\xi}_{\Lambda}$ is the polarisation of the $\Lambda$ hyperon
in the $\Lambda$ rest frame obtained by a Lorentz transformation
along $\vec{n}_\Lambda$ from the $\Lambda_c^+$ baryon rest
frame.

Note that if time-reversal invariance is valid and final-state interactions are ignored, then the parameter $\beta_{\Lambda_c} =0$.

The spin direction of $\Lambda$ baryon could be determined by measuring the proton-decay asymmetry in the $\Lambda$ rest frame through the relation
\begin{equation}\label{eq:0020}
\frac{1}{N}\frac{d\,N}{d\,\Omega}=\frac{1}{4\pi}\left(1+\alpha_\Lambda
\vec{\xi}_{\Lambda}\cdot \hat{\vec{p}} \right),
\end{equation}
where $\hat{\vec{p}}$ is a unit vector along the daughter-proton direction and $\vec{\xi}_{\Lambda}$ is given by Eq.(\ref{eq:0018}).
If parameters $\alpha_{\Lambda_c}$, $\beta_{\Lambda_c}$, and $\gamma_{\Lambda_c}$ are known,
by projection of Eq.~(\ref{eq:0018}) on three orthogonal axes, one can find the components of
the polarisation vector of $\vec{\xi}_{\Lambda_c}$ for each event.
Thus, in this way all the information about polarisation $\Lambda_c^+$
can be obtained from the decay of $\Lambda$ baryon, without the need to
refer to asymmetries or distributions in the rest frame of the $\Lambda_c^+$.
Note that methods based on relation between polarisation of parent baryon and daughter baryon
have been applied in studies of hyperon decays (see, e.g., Refs.~\cite{Handler:1982gd, Aston:1985sn}).

Then using the very well measured value of $\alpha_\Lambda=0.642\pm0.013$~\cite{Tanabashi:2018oca} we could
achieve to obtain $\alpha_{\Lambda_c}$ and $\xi_{\Lambda_c}$  separately, for example, using Eqs.~(\ref{eq:0010}) and
(\ref{eq:0012}). So far $\alpha_{\Lambda_c}$ is measured with less precision, 
$\alpha_{\Lambda_c}=-0.91\pm 0.15$~\cite{Tanabashi:2018oca}. Measuring $\alpha_{\Lambda_c}$ and $\xi_{\Lambda_c}$ 
with much higher statistics data of LHCb  will be very interesting in the future. In particular, in view of results of $\Lambda_b$ polarisation
measurement at LHCb~\cite{Aaij:2013oxa}, $\xi_b = 0.06 \pm 0.07 \pm 0.02, \, \alpha_b=0.05 \pm 0.17 \pm 0.07$,  and at
CMS~\cite{Sirunyan:2018bfd}, $\xi_b = 0.00 \pm 0.06 \pm 0.06, \, \alpha_b=0.14 \pm 0.14 \pm 0.10$, which show that $\Lambda_b$ is
little polarised, it is most important to measure the
$\Lambda_c$ polarisation.  In the case of $\Lambda_c$, the large value of $\alpha_{\Lambda_c} $ would
help to measure both $\alpha_{\Lambda_c}$ and $\xi_{\Lambda_c}$ at a much higher precision.

\subsection{The case of $\Lambda_c \to  p \, K \, \pi$}
The use of the  $\Lambda_c \to p \, K \, \pi$ decays is also interesting because of  its  largest branching fraction, $6.23\pm 0.33$ \%. E791 experiment \cite{Aitala:1999uq} studied three main intermediate states:  
$\Lambda_c\to [{K}^*(890)\,p, \,  \Delta^{++} (1232) \, K, \,  \Lambda(1520) \, \pi]\to p \,  K \, \pi$.
In this analysis, the $\Lambda_c \to [{K}^*(890) \, p$ ($\Delta^{++}(1232) \, K, \,  \Lambda(1520) \, \pi] \to p \,  K \, \pi$ decay is parametrized by  4 (2) complex helicity amplitudes.  
Those are given as 8 (4) real parameters $(E_{1\sim 4}, \, \phi_{E_{1\sim 4}})$ 
for ${K}^*(890) \, p$ channel, $(F_{1\sim 2}, \, \phi_{F_{1\sim 2}})$ for $\Delta^{++}(1232) \, K$ channel 
and $(H_{1\sim 2}, \, \phi_{H_{1\sim 2}})$ for $\Lambda(1520) \, \pi$ channel. 
Furthermore, the continuum background is modeled by the $S$-wave amplitude which introduce another 8 real parameters. 
Including the polarisation parameter $\xi_{\Lambda_c}$ (denoted as ${\bf P}$ in the paper \cite{Aitala:1999uq}), a total of 25 parameters are fitted by using the full angular and Dalitz variables.

From the amplitude parameters, we can also obtain $\alpha$ for each resonance
\begin{equation}
\alpha_{K^*p} = 0.66\pm 0.28, \quad
\alpha_{\Delta^{++}K} = -0.67\pm 0.30, \quad
\alpha_{\Lambda(1520)\pi} = -0.11\pm 0.60.
\label{eq:alphas}
\end{equation}
Note that we find a different value for $\alpha_{K^*p}$ with respect to \cite{Botella:2016ksl}.
The higher values of $\alpha_{K^*p}$ and $\alpha_{\Delta^{++}K}$ make the use of these channels interesting for polarisation studies, though the error is still too large to be able to conclude.

It would be interesting to repeat this analysis at LHCb, which has much higher rate of the $\Lambda_c^+$ production. 
The crucial point of this measurement lies on the value of the polarisation of $\Lambda_c^+$ produced at LHCb.

\section{Conclusions}

Recently a new experiment  for measuring the magnetic moment of the $\Lambda_c$ baryon using a bent crystal is proposed~\cite{Bezshyyko:2017var,Botella:2016ksl}. 
Although the magnetic moment of charm quark is a fundamental property, which enters to various QCD computation, it has never been determined precisely. 
This experimental proposal can provide us its very first measurement.

The theoretical predictions of the magnetic moment of charmed baryons suffer from the hadronic uncertainties. On the other hand, in the quark model, the $\Lambda_c$ magnetic moment is equal to the charm quark magnetic moment and this can remain correct by including the light degree of freedom, due to the spin structure of the light degree of freedom. This makes $\Lambda_c$ to be the most simple object to study the charm magnetic moment, up to the charm quark mass uncertainty.

We have introduced relations among magnetic moments of different charmed baryons, which could cancel the charm quark mass ambiguity. We have also related  the $\Lambda_c$ magnetic moment to 
the charm quark magnetic moment measurement by the radiative quarkonium decays, using angular distribution of successive $\psi(2S)\to \chi_{cJ}\gamma\to J/\psi \gamma$ decays, which were performed by the CLEO and the BESIII collaborations. Interestingly, we observe a slight tension: the obtained value is  higher than most of the theoretical predictions of the $\Lambda_c$ magnetic moment. Further improvement of quarkonium radiative decay is very important.

It has been shown that when measuring the $g$-factor of $\Lambda_c$ directly, ${\rm i.e.}$ through spin precession, the knowledge of weak decay parameter $\alpha$ and initial polarisation $\xi$ could reduce the data taking time needed to reach the error of $\Delta g=0.1$. 
The $\alpha$ parameter can be pre-measured in another experiment which has the same experimental setting (${\rm i.e.}$ $p_{\rm t}$ and $\xi$), ${\rm e.g.}$ by SMOG experiment, though the statistics are limited. 
Alternatively, we may use the very high statistic data of LHCb to extract separately $\alpha$ and $\xi$ values and we can extrapolate the $\xi$ to the required $p_{\rm t}$ range by using theory.
The error on $g$-factor at a few percents could be reached after reconstructing $10^4$ decays of deflected $\Lambda_c$ baryons, and in this case it is more efficient to measure $g$-factor and $\alpha\,\xi$ simultaneously.

We estimated the error on $g$-factor using these two approaches and compared the measurement efficiencies at two places: at LHCb detector and at momentum cleaning area of LHC (IR3), proposed in~\cite{Mirarchi:2019ft}.
The latter case requires building a new dedicated detector but it would need about
7.5 times less data taking time in order to reach the same precision.

We found a special orientation of the crystal that gives the opportunity to measure the $\Lambda_c$ dimensionless electric dipole moment almost with the same precision as its $g$-factor.
Our calculations show that this method is about $40$ times more efficient in terms of data taking time
with respect to the one proposed in \cite{Bagli:2017foe}.

The estimated error on $g$-factor after 10 years of data taking using the setup of $40\,$mm tungsten target and germanium crystal at LHCb and IR3 is $\Delta g=0.100$ and $\Delta g=0.037$, respectively.
With a slight adjustment of the crystal orientation (rotating the crystal by a few milliradians) the $\Lambda_c$ EDM could be measured with an error $\Delta d = 2.6 \times 10^{-16} e\,$cm at LHCb and $\Delta d = 1.0 \times 10^{-16} e\,$cm at IR3.

\section*{Acknowledgments} 

This research was partially conducted in the scope of the IDEATE International Associated
Laboratory (LIA).
A.Yu.K., V.A.K. and A.S.F. acknowledge partial support by 
the National Academy of Sciences of Ukraine (project KPKVK 6541230, and project No. Ts-3/53-2019)
and the Ministry of Education and Science of Ukraine (project No. 0117U004866).

\section*{Appendix A: Quark model relations } 
\label{sec:app_a}

In this Appendix we summarise expressions for the magnetic dipole moments (MDM) of the single and double charmed baryons in non-relativistic 
constituent quark model.   Only baryons with $J^P=\frac{1}{2}^+$ are  considered here. Some properties of these baryons are shown in 
Table~\ref{tab:charmed_baryons}.  For a review of the charm baryons see Ref.~\cite{Korner:1991ap}.  

\begin{table}[tbh]
\caption{Properties of the single and double charmed baryons.  The antisymmetric and symmetric in flavour 
functions are defined as  $[q_1 \, q_2] \equiv \frac{1}{\sqrt{ 2}}(q_1 q_2 - q_2 q_1) $ 
and $\{ q_1 \, q_2 \} \equiv \frac{1}{\sqrt{2}}(q_1 q_2 + q_2 q_1) $, respectively.
The production cross section of baryon
at the LHC fixed-target mode ($\sqrt{s}\approx110\,$GeV) and in collider conditions
($\sqrt{s}= 13\,$TeV) -- results of Pythia simulation.
}
\begin{center}
\begin{tabular}{| l |c|c|c|c|c|c|c|c|c|}
  \hline
~Baryon~&~Flavor~&$~SU(3)_f$~&~~$I$~~~&$~I_z~$&~ Charm ~&~ Mass (MeV) ~&\multicolumn{2}{c|}{~Cross section ($\mu$barn)~}&~Life-length~ \\
          &~content~&      & & &         &\cite{Tanabashi:2018oca}&~fixed target~&collider                       &~or decay width ~\\
\hline
~~~$\Lambda_c^+$  &$  [ud]c $&$\bar{3}$&          0        &              0      &1&$2286.5\pm0.1$&10.13&758.1&$60.0\pm1.2\,\mu$m\\
~~~$\Xi_c^+ $           &$  [us]c $&$\bar{3}$&$\frac{1}{2}$&$+\frac{1}{2}$&1&$2467.9\pm0.2$&0.588&65.5&$132.5\pm7.8\,\mu$m\\
~~~$\Xi_c^0$            &$  [ds]c $&$\bar{3}$&$\frac{1}{2}$&$ -\frac{1}{2}$&1&$2470.9\pm0.3$&0.510&65.6&$33.6\pm3.6\,\mu$m\\
~~~$\Sigma_c^{++}$ &$  uuc  $&$      {6}$&           1       &$        +1     $&1&$2454.0\pm0.1$&0.863&42.0&$1.9\pm0.1\,$MeV\\
~~~$\Sigma_c^{+}$   &$\{ud\}c$&$      {6}$&           1      &$           0     $&1&$2452.9\pm0.4$&0.697&42.2&$<4.6\,$MeV\\
~~~$\Sigma_c^{0}$   &$  ddc  $&$      {6}$&           1       &$         -1     $&1&$2453.8\pm0.1$&0.461&41.6&$1.8\pm0.1\,$MeV\\
~~~$\Xi_c^{\prime+}$&$\{us\}c$&$       6 $&$\frac{1}{2}$&$+\frac{1}{2}$&1&$2578.4\pm0.5$&0.083&6.3&--\\
~~~$\Xi_c^{\prime0}$&$\{ds\}c$&$       6 $&$\frac{1}{2}$&$ -\frac{1}{2}$&1&$2579.2\pm0.5$&0.072&6.6&--\\
~~~$\Omega_c^{0}$  &$  ssc  $&$       6 $&$        0      $&$          0      $&1&$2695.2\pm1.7$&0.028&3.0&$80.3\pm10\,\mu$m\\
~~~$\Xi_{cc}^{++}$    &$  ccu  $&$       3 $&$\frac{1}{2}$&$+\frac{1}{2}$&2&$3621.4\pm0.8$&$<10^{-4}$&$\sim10^{-3}$&$76.7\pm10\,\mu$m\\
~~~$\Xi_{cc}^{+}$      &$  ccd  $&$       3 $&$\frac{1}{2}$&$ -\frac{1}{2}$&2&$3518.9\pm0.9$&$<10^{-4}$&$<10^{-3}$&--\\   
~~~$\Omega_{cc}^{+}$&$ccs  $&$       3 $&$        0     $&$           0     $&2&         --               &$<10^{-4}$&$\sim10^{-3}$&--\\          
\hline
\end{tabular}
\end{center}
\label{tab:charmed_baryons}
\end{table}

In the 2nd column of Table~\ref{tab:charmed_baryons} the flavour wave functions of baryons are shown.   
To construct spin-flavour wave functions of the baryons with total spin $J =\tfrac{1}{2}$ and 
its projection  $J_z =  +\tfrac{1}{2}$ the flavour  
functions are to be combined with either antisymmetric spin function 
\begin{equation}
\psi_{asym} = \frac{1}{\sqrt{2}}( \, \uparrow \downarrow \uparrow - \downarrow \uparrow \uparrow \, )  ,
\label{eq:A01}
\end{equation} 
or symmetric one   
\begin{equation}
\psi_{sym} = \frac{1}{\sqrt{6}} 
[ \, 2 \uparrow \uparrow \downarrow - (\downarrow \uparrow+\uparrow \downarrow)  \uparrow \, ],
\label{eq:A02}
\end{equation} 
with respect to interchange of particles 1 and 2.

The magnetic dipole moment of baryon $B$ is calculated from the definition
\begin{equation}
\mu_B = \langle   B;  \tfrac{1}{2},  +\tfrac{1}{2}  | \, \mu_1 \sigma_{1 z} + \mu_2 \sigma_{2 z}  + \mu_3 \sigma_{3 z}
 \,  | B; \, \tfrac{1}{2}, +\tfrac{1}{2}  \rangle ,
\label{eq:A06}
\end{equation}
where $\mu_i = \tfrac{g_i}{2} \tfrac{e Q_i}{2m_i} $ is the magnetic moment of the  i-th quark.    

Below we list wave functions of the baryons from $SU(3)_f$ anti-triplet from Table~\ref{tab:charmed_baryons}: 
\begin{eqnarray}
&& | \Lambda_c^+; \tfrac{1}{2},  +\tfrac{1}{2} \rangle 
= \frac{1}{2} ( u_{\uparrow} d_{\downarrow}  c_{\uparrow}  -u_{\downarrow} d_{\uparrow}  c_{\uparrow} 
-d_{\uparrow} u_{\downarrow }  c_{\uparrow} +d_{\downarrow} u_{\uparrow } c_{\uparrow}  ), \\
\label{eq:A03a}
&& | \Xi_c^+; \tfrac{1}{2},  +\tfrac{1}{2} \rangle 
= \frac{1}{2 } ( u_{\uparrow} s_{\downarrow}  c_{\uparrow}  -u_{\downarrow} s_{\uparrow}  c_{\uparrow} 
-s_{\uparrow} u_{\downarrow }  c_{\uparrow} +s_{\downarrow} u_{\uparrow } c_{\uparrow} ), \\
\label{eq:A03b}
&& | \Xi_c^0; \tfrac{1}{2},  +\tfrac{1}{2} \rangle 
= \frac{1}{2 } ( d_{\uparrow} s_{\downarrow}  c_{\uparrow}  -d_{\downarrow} s_{\uparrow}  c_{\uparrow} 
-s_{\uparrow} d_{\downarrow }  c_{\uparrow} +s_{\downarrow} d_{\uparrow } c_{\uparrow}) . 
\label{eq:A03c}
\end{eqnarray}

Wave functions for the $SU(3)_f$ sextet read
\begin{eqnarray}
 && |\Sigma_c^{++}; \tfrac{1}{2},  +\tfrac{1}{2} \rangle 
 = \frac{1}{\sqrt{6}} 
( 2 u_{\uparrow} u_{\uparrow} c_\downarrow - u_{\uparrow} u_{\downarrow} c_\uparrow
-u_{\downarrow} u_{\uparrow} c_\uparrow ) , \\
\label{eq:A04a}
 && |\Sigma_c^+; \tfrac{1}{2},  +\tfrac{1}{2} \rangle 
 = \frac{1}{2 \sqrt{3}} 
( 2 u_{\uparrow} d_{\uparrow} c_\downarrow + 2 d_{\uparrow} u_{\uparrow} c_\downarrow
-u_{\uparrow} d_{\downarrow} c_\uparrow  - d_{\uparrow} u_{\downarrow}  c_{\uparrow}
- u_{\downarrow} d_{\uparrow } c_\uparrow - d_{\downarrow} u_{\uparrow } c_\uparrow ), \\
\label{eq:A04b}
&& |\Sigma_c^0; \tfrac{1}{2},  +\tfrac{1}{2} \rangle 
 = \frac{1}{\sqrt{6}} 
( 2 d_{\uparrow} d_{\uparrow} c_\downarrow - d_{\uparrow} d_{\downarrow} c_\uparrow
-d_{\downarrow} d_{\uparrow} c_\uparrow ) , 
\label{eq:A04c}
\end{eqnarray}
and
\begin{eqnarray}
&& |\Xi_c^{\prime +}; \tfrac{1}{2},  +\tfrac{1}{2} \rangle 
 = \frac{1}{2 \sqrt{3}} 
( 2 u_{\uparrow} s_{\uparrow} c_\downarrow + 2 s_{\uparrow} u_{\uparrow} c_\downarrow
-u_{\uparrow} s_{\downarrow} c_\uparrow  - s_{\uparrow} u_{\downarrow}  c_{\uparrow}
- u_{\downarrow} s_{\uparrow } c_\uparrow - s_{\downarrow} u_{\uparrow } c_\uparrow ), \\
\label{eq:A05a}
&& |\Xi_c^{\prime 0}; \tfrac{1}{2},  +\tfrac{1}{2} \rangle 
 = \frac{1}{2 \sqrt{3}} 
( 2 d_{\uparrow} s_{\uparrow} c_\downarrow + 2 s_{\uparrow} d_{\uparrow} c_\downarrow
-d_{\uparrow} s_{\downarrow} c_\uparrow  - s_{\uparrow} d_{\downarrow}  c_{\uparrow}
- d_{\downarrow} s_{\uparrow } c_\uparrow - s_{\downarrow} d_{\uparrow } c_\uparrow ), \\
\label{eq:A05b}
&& |\Omega_c^0; \tfrac{1}{2},  +\tfrac{1}{2} \rangle 
 = \frac{1}{\sqrt{6}} 
( 2 s_{\uparrow} s_{\uparrow} c_\downarrow - s_{\uparrow} s_{\downarrow} c_\uparrow
-s_{\downarrow} s_{\uparrow} c_\uparrow  ).
 \label{eq:A05c}
\end{eqnarray}

Finally, the double-charmed baryons from $SU(3)_f$ triplet have wave functions
\begin{eqnarray}
 && |\Xi_{cc}^{++}; \tfrac{1}{2},  +\tfrac{1}{2} \rangle  = \frac{1}{\sqrt{6}} 
( 2 c_{\uparrow} c_{\uparrow} u_\downarrow - c_{\uparrow} c_{\downarrow} u_\uparrow
-c_{\downarrow} c_{\uparrow} u_\uparrow   ), \\
\label{eq:A06a}
&& |\Xi_{cc}^+; \tfrac{1}{2},  +\tfrac{1}{2} \rangle = \frac{1}{ \sqrt{6}} 
( 2 c_{\uparrow} c_{\uparrow} d_\downarrow - c_{\uparrow} c_{\downarrow} d_\uparrow
-c_{\downarrow} c_{\uparrow} d_\uparrow ), \\
\label{eq:A06b}
&& |\Omega_{cc}^+; \tfrac{1}{2},  +\tfrac{1}{2} \rangle = \frac{1}{\sqrt{6}} 
( 2 c_{\uparrow} c_{\uparrow} s_\downarrow - c_{\uparrow} c_{\downarrow} s_\uparrow
-c_{\downarrow} c_{\uparrow} s_\uparrow ).
\label{eq:A06c}
\end{eqnarray}
These wave functions are normalized to unity. 

The magnetic moments of the charmed baryons are shown in Table \ref{tab:magnetic}.
\begin{table}[tbh]
\caption{ MDM of charmed baryons in terms of MDM of constituent quarks. In the 3d column 
$\theta_+$ is mixing angle for $\Xi_c^+$ and $\Xi_c^{\prime +}$,  
and $\theta_0$ is mixing angle for $\Xi_c^0$ and $\Xi_c^{\prime 0}$, and `n.m.' stands for `not modified'.}
\begin{center}
\begin{tabular}{|l |l | l |}
  \hline
Baryon  & MDM   & MDM with mixing  \\
	\hline
  $\Lambda_c^+$&   $\mu_c$ &  {\rm n.m.} \\
$\Xi_c^+ $ &   $\mu_c$ &  $  \mu_c \cos^2 \theta_++  \frac{1}{3} (2 \mu_u +2 \mu_s -\mu_c) \sin^2 \theta_+ +  \frac{1}{\sqrt{3}}(\mu_s -\mu_u) \sin 2\theta_+$ \\
$\Xi_c^0$  &   $\mu_c$ & $  \mu_c \cos^2 \theta_0 +  \frac{1}{3} (2 \mu_d +2 \mu_s -\mu_c) \sin^2 \theta_0 +  \frac{1}{\sqrt{3}}(\mu_s -\mu_d) \sin 2\theta_0$ \\
$\Sigma_c^{++} $ &    $ \frac{1}{3}(4 \mu_u - \mu_c)$ & {\rm n.m.}   \\
$\Sigma_c^{+} $ &   $ \frac{1}{3}(2 \mu_u +2 \mu_d - \mu_c)$  & {\rm n.m.} \\
$\Sigma_c^{0} $ &    $ \frac{1}{3}(4 \mu_d - \mu_c)$ & {\rm n.m.} \\
$\Xi_c^{\prime +}$  &  $\frac{1}{3}(2 \mu_u +2 \mu_s - \mu_c)$ & $  \mu_c \sin^2 \theta_+ 
+  \frac{1}{3} (2 \mu_u +2 \mu_s -\mu_c)\cos^2 \theta_+ - \frac{1}{\sqrt{3}}(\mu_s -\mu_u)\sin 2\theta_+ $\\
$\Xi_c^{\prime 0}$  &   $\frac{1}{3}(2 \mu_d +2 \mu_s - \mu_c)$  & $  \mu_c \sin^2 \theta_0 
+  \frac{1}{3} (2 \mu_d +2 \mu_s -\mu_c) \cos^2 \theta_0 -  \frac{1}{\sqrt{3}}(\mu_s -\mu_d) \sin 2\theta_0$\\
$\Omega_c^{0}$  &   $\frac{1}{3}(4 \mu_s  - \mu_c)$  & {\rm n.m.} \\
$\Xi_{cc}^{++}$  &   $\frac{1}{3}(4 \mu_c  - \mu_u)$  & {\rm n.m.} \\  
$\Xi_{cc}^{+}$  &  $\frac{1}{3}(4 \mu_c - \mu_d)$ & {\rm n.m.} \\   
$\Omega_{cc}^{+}$  &  $\frac{1}{3}(4\mu_c - \mu_s)$  & {\rm n.m.} \\          
\hline
\end{tabular}
\end{center}
\label{tab:magnetic}
\end{table}
 
Important modification included in Table~\ref{tab:magnetic} is the effect of mixing which 
was first addressed in \cite{Franklin:1981rc} and studied in detail in Refs.~\cite{Bernotas:2013, Simonis:2018rld}. 
The mixing appears between the states $\Xi_c^+$ and $\Xi_c^{\prime +}$, 
and between the  states $\Xi_c^0$ and $\Xi_c^{\prime 0}$.  According to \cite{Bernotas:2013},
 the mixing is of little importance for the neutral baryons 
$\Xi_c^0$ and $\Xi_c^{\prime 0}$, while it is essential for the charged ones 
$\Xi_c^+$ and $\Xi_c^{\prime +}$.  This is related to different magnitude of the transition operators 
in Table~\ref{tab:magnetic}, namely $ \tfrac{1}{\sqrt{3}}|\mu_s - \mu_u| \gg \tfrac{1}{\sqrt{3}}|\mu_s - \mu_d|$. 
The transition magnetic moments between $\Xi_c^+$ and $\Xi_c^{\prime +}$, and between  
$\Xi_c^0$ and $\Xi_c^{\prime 0}$ are
\begin{eqnarray}
&& \mu_{\, \Xi_c^{\prime +} \to \Xi_c^{+}} =  \frac{1}{\sqrt{3}} (\mu_s - \mu_u) \cos 2 \theta_+
+ \frac{1}{3} (\mu_u + \mu_s - \mu_c)  \sin 2 \theta_+ , 
\label{eq:A08a}   \\
, \qquad 
&& \mu_{\, \Xi_c^{\prime 0} \to \Xi_c^{0}} = \frac{1}{\sqrt{3}} (\mu_s - \mu_d) \cos 2 \theta_0 
+  \frac{1}{3} (\mu_d + \mu_s - \mu_c) \sin 2 \theta_0, 
\label{eq:A08b}
\end{eqnarray}
and one also finds that $ |\mu_{\, \Xi_c^{\prime +} \to \Xi_c^{+}}| \gg |\mu_{\, \Xi_c^{\prime 0} \to \Xi_c^{0}}|$.  
The mixing for the baryons $\Xi_c^+$ and $\Xi_c^{\prime +}$    
may complicate interpretation of  $\Xi_c^+$ MDM as being entirely due to the charm quark.

\newpage


\end{document}